\documentclass[preprint,preprintnumbers,amsmath,amssymb]{revtex4}

\usepackage{graphicx}
\usepackage{dcolumn}
\usepackage{bm}
\usepackage{rotating}

\begin{document}

\title[Unified Approach for Nonlinear Oscillators ]{A Simple and Unified 
Approach to Identify Integrable Nonlinear Oscillators and Systems}

\author{V. K. Chandrasekar$^{\dag}$, S. N. Pandey$^{\dag\dag}$ 
M. Senthilvelan$^{\dag}$ and  M. Lakshmanan$^{\dag}$}
\affiliation {$^{\dag}$ Centre for Nonlinear Dynamics, Department of Physics, 
Bharathidasan University, Tiruchirappalli - 620 024, India }
\affiliation {$^{\dag\dag}$ Department of Physics, Motilal Nehru National Institute of 
Technology, Allahabad-211 004, India}

\date{\today} 

\begin{abstract}
In this paper, we consider a generalized second order nonlinear ordinary
differential equation of the
form
$\ddot{x}+(k_1x^q+k_2)\dot{x}+k_3x^{2q+1}+k_4x^{q+1}+\lambda_1x=0$, where
$k_i$'s, $i=1,2,3,4$, $\lambda_1$ and $q$ are arbitrary parameters, which
includes 
several physically important nonlinear oscillators such as the simple 
harmonic oscillator, 
anharmonic oscillator, force-free Helmholtz oscillator, force-free Duffing and
Duffing-van der Pol 
oscillators, modified Emden type equation and its hierarchy, generalized 
Duffing-van der Pol oscillator equation hierarchy and so on and investigate the
integrability properties of this rather general equation. 
We identify several new integrable cases for arbitrary value of the exponent
$q,\;q\in R$. The $q=1$ and $q=2$ cases are analyzed in detail and the results
are generalized to arbitrary $q$. Our results show that
many classical integrable nonlinear oscillators can be derived as sub-cases of
our results and significantly enlarge the list of integrable equations that 
exist in the contemporary
literature. To explore the above underlying results we use the recently
introduced generalized extended Prelle-Singer procedure applicable to second 
order ODEs. As
an added advantage of the method we not only identify integrable regimes but
also construct integrating factors, integrals of motion and general solutions
for the integrable cases, wherever possible, and bring out the mathematical
structures associated with each of the integrable cases.   
\end{abstract}

\maketitle
\section{Introduction}
\label{sec1}
\subsection{Overview of the problem}
\label{sec11}
In a recent paper$\footnotesize^1$ we have shown that the force-free Duffing-van der Pol
(DVP) oscillator,
\begin{eqnarray} 
\ddot{x}+(\alpha+\beta x^2)\dot{x}-\gamma x +x^3=0, \label {int01}
\end{eqnarray}
is integrable for the parametric restriction $\alpha=\frac {4}{\beta}$ and
$\gamma=-\frac{3}{\beta^2}$. In Eq.~(\ref{int01}) over dot denotes
differentiation with respect to $t$ and $\alpha,\;\beta$ and $\gamma$ are
arbitrary parameters. Under the transformation 
\begin{eqnarray} 
w= -x e^{\frac{1}{\beta}t},\;\;\;\;\; 
z= e^{-\frac{2}{\beta}t},\label {int02}
\end{eqnarray}
Eq.~(\ref{int01}) with restriction $\alpha=\frac {4}{\beta}$ and
$\gamma=-\frac{3}{\beta^2}$ was shown to be transformable to the form 
\begin{eqnarray} 
w''-\frac{\beta^2}{2}w^2w'=0,\label {int02a}
\end{eqnarray}
which can then be integrated$\footnotesize^1$. 

In a parallel direction, while performing the invariance analysis of a 
similar
kind of problem, we find that not only the Eq.~(\ref{int01}) but also its
generalized version,
\begin{eqnarray} 
\ddot{x}+(\frac {4}{\beta}+\beta x^2)\dot{x}+\frac{3}{\beta^2} x 
+x^3+\delta x^5=0,\quad \delta= \mbox{arbitrary parameter}, 
\label {int03}
\end{eqnarray}
is invariant under the same set of Lie point symmetries$\footnotesize^2$.
As a consequence one can use the same transformation
(\ref{int02}) to integrate the Eq.~(\ref{int03}). 
The transformation (\ref{int02}) modifies Eq.~(\ref{int03}) to the form
\begin{eqnarray} 
w''-\frac {\beta^2}{2}w^2w'+\delta w^5=0
\label {int04}
\end{eqnarray}
which is not so simple to integrate straightforwardly. However, we observe that 
this equation coincides with the second equation in the so called modified Emden
equation (MEE) hierarchy, investigated
by Feix et al.$\footnotesize^3$,
\begin{eqnarray} 
\ddot{x}+x^l\dot{x}+gx^{2l+1}=0,\;\;\;l=1,2,\ldots, n,
\label {int05}
\end{eqnarray}
where $g$ is an arbitrary parameter.
 
In fact Feix et al.$\footnotesize^3$ have shown that through a direct
transformation to a third order equation the above Eq.~(\ref{int05}) can be integrated
to obtain the general solution for the specific choice of the parameter $g$, 
namely, for $g=\frac{1}{(l+2)^2}$. For this choice of $g$, the general 
solution of (\ref{int05}) can be written as 
\begin{eqnarray}            
x(t)=\bigg(\frac{(2+3l+l^2)(t+I_1)^l}{l(t+I_1)^{l+1}
+(2+3l+l^2)I_2}\bigg)^{\frac{1}{l}}, 
\;\; I_1, I_2:\;\mbox{arbitrary constants}. \label {hie05}
\end{eqnarray}

Consequently Eq.~(\ref{int03}) can be integrated under
the specific parametric choice $\delta=\frac{1}{16}$, and it belongs to the 
$l=2$ case of the MEE hierarchy (\ref{int05}) with $g=\frac{1}{16}$. Now the
question arises as to whether there exist other new integrable second order
nonlinear differential equations which 
are linear in $\dot{x}$ and containing 
fifth and other powers of nonlinearity. As far as our knowledge goes only few equations in
this class have been shown to be integrable. 
For example, Smith$\footnotesize^4$ had investigated a class of 
nonlinear equations coming under the category
\begin{eqnarray}            
\ddot{x}+f(x)\dot{x}+g(x)=0,
\label {int07a}
\end{eqnarray}
with $f(x)=(n+2)bx^n-2a$ and $g(x)=x(c+(bx^n-a)^2)$ where $a,b,c$ and $n$ are
arbitrary parameters. He had shown that 
the Eq.~(\ref{int07a}) with this specific forms of $f$ and $g$ admits explicit
oscillatory solutions. However,
one can also expect that there should be a number of integrable equations which also 
admit solutions which are both oscillatory and non-oscillatory types in the 
class
\begin{eqnarray}            
\ddot{x}+(k_1x^q+k_2)\dot{x}+k_3x^{2q+1}+k_4x^{q+1}+\lambda_1x=0,\qquad q\in R,
\label {int08}
\end{eqnarray}
where $k_i$'s, $i=1,2,3,4$ and $\lambda_1$ are arbitrary parameters.  
When $q=1$, Eq.~(\ref{int08}) becomes the generalized MEE
\begin{eqnarray}            
\ddot{x}+(k_1x+k_2)\dot{x}+k_3x^3+k_4x^2+\lambda_1x=0,
\label {int07}
\end{eqnarray}
and for $q=2$ it becomes
\begin{eqnarray}            
\ddot{x}+(k_1x^2+k_2)\dot{x}+k_3x^5+k_4x^3+\lambda_1x=0.
\label {int07b}
\end{eqnarray}
We note that Eq.~(\ref{int03}) is a special case of (\ref{int07b}). 

Needless to say Eq.~(\ref{int08}) is a unified model for several ground
breaking physical systems which includes simple harmonic oscillator, 
anharmonic oscillator, force-free Helmholtz oscillator, force-free Duffing
oscillator, MEE hierarchy, generalized DVP hierarchy and so on.

As noted earlier there exists no rigorous mathematical analysis in the
literature for the second order nonlinear differential equations which contain 
fifth or higher degree nonlinearity in $x$ 
and linear in $\dot{x}$ and the results are very scarce on integrability or 
exact solutions. Our motivation to 
analyze this problem is not only to explore new integrable cases/systems of
Eq.~(\ref{int08}) but 
also to synthesize all earlier results under one approach.

Having described the problem and motivation now we can start analyzing the
integrability properties of Eq.~(\ref{int08}). To identify 
the integrable regimes  we employ the recently 
introduced extended Prelle-Singer procedure applicable
to second order ODEs$\footnotesize^{5-11}$. Through this
method we not only identify integrable regimes but also construct integrating
factors, integrals of motion and general solution for the integrable cases, 
wherever possible.
\subsection{Results}
\label{sec12}
We unearth several new integrable equations for any real value of the exponent 
$q$ in Eq.~(\ref{int08}).
In the following we summarize the results for the case 
$\it{q=\;arbitrary}$ only and discuss in detail the $q=1,\;q=2$ and 
$q=\;arbitrary$ cases separately in the following sections.

For the choice $\it{q=\;arbitrary}$ we find that the following equations are
completely integrable (after suitable reparametrizations), all of which appear 
to be new to the literature:
\begin{eqnarray}
&&\ddot{x}+(k_1x^q+(q+2)k_2)\dot{x}
+k_1k_2x^{q+1}+(q+1)k_2^2x=0
\label {ncom11a}\\
&& \ddot{x}+((q+2)k_1x^q+k_2)\dot{x}+k_1^2x^{2q+1}
+k_1k_2x^{q+1}+\lambda_1x=0
\label {ncom7}\\
&&\ddot{x}+(q+4)k_2\dot{x}+k_4x^{q+1}+2(q+2)k_2^2x=0
\label {ncom11}\\
&&\ddot{x}+((q+1)k_1x^q+k_2)\dot{x}+\frac{(r-1)}{r^2}[(q+1)
k_1^2x^{2q+1}
\nonumber\\
&&\qquad\qquad\qquad\qquad\qquad+(q+2)k_1k_2x^{q+1}+k_2^2x]=0,\;\;r\neq0
\label {ncom9}\\
&&\ddot{x}+((q+1)k_1x^q+(q+2)k_2)\dot{x}+(q+1)
[\frac{(r-1)}{r^2}k_1^2x^{2q+1}\nonumber\\
&&\qquad \qquad \qquad\qquad\qquad\qquad+k_1k_2x^{q+1}+k_2^2x]=0,\;\;r\neq0,
\label {ncom6c}
\end{eqnarray}
where $k_1,\;k_2,\;k_4,\;\lambda_1$ and $r$ are 
arbitrary parameters. 
We stress that the above results are true for any arbitrary values of $q$. 
We discuss the special cases, namely, $q=1$ and 
$q=2$ separately in detail in sections 3 and 4 in order to put the results of
$q$ arbitrary case in proper perspective. 

We show that the Eq.~(\ref{ncom11a}) is nothing but a generalization of the 
Duffing-van der Pol oscillator Eq.~(\ref{int01}). In a recent 
work$\footnotesize^{1,9}$ three of the present authors have
established the integrability of Eq.~(\ref{ncom11a}) with $q=2$. However,
in this work we show that the generalized Eq.~(\ref{ncom11a}) itself is 
integrable. Eq.~(\ref{ncom7}) is nothing but the generalized MEE among which the 
hierarchy of
Eq.~(\ref{int05}), studied by Feix et al.$\footnotesize^3$, can be identified 
as a sub-case.
In fact the general solution constructed by Feix et al., Eq.~(\ref{hie05}), can be
derived straightforwardly as a sub-case. Eq.~(\ref{ncom7}) also
contains the family of equations studied by Smith$\footnotesize^4$. 
In particular the
latter author have derived general solution for the case $k_2^2<4\lambda_1$, 
which turns out to be an oscillatory one. However, in this work, we show that 
even for
arbitrary values of $k_2$ and $\lambda_1$ one can construct the general
solution. Interestingly, the system (\ref{ncom11})
generalizes several physically important nonlinear oscillators. For example, in
the case $q=1$ and $2$, Eq.~(\ref{ncom11}) provides us the force-free Helmholtz 
and Duffing
oscillators, respectively, whose nonlinear dynamics is well documented in the
literature$\footnotesize^{12-16}$. Here, we present certain 
integrable generalizations of these nonlinear oscillators.
Eq.~(\ref{ncom9}) admits a conservative Hamiltonian for all
values of the parameters $r,\;k_1$ and $k_2$ and any integer value of $q$. 
We also provide the explicit form of the Hamiltonian for all
values of $q$. As a result we conclude that it is a Liouville integrable system.
As far as Eq.~(\ref{ncom6c}) is concerned we construct a time dependent 
integral of motion and transform the latter to time independent Hamiltonian 
one and thereby ensuring its Liouville integrability.

The plan of the paper is us follows. In the following section we briefly
describe the extended Prelle-Singer procedure applicable to second order ODEs. 
In Sec.~\ref{sec3}, we
consider the case $q=1$ in (\ref{int08}) and identify the integrable parametric 
choices of
this equation through the extended PS procedure. To do so first we identify the
integrable cases where the system admits time independent integrals and
construct explicit conservative Hamiltonians for the respective parametric
choices. We then identify the cases which
admit explicit time dependent integrals of motion. To establish the complete
integrability of these cases we use our own procedure and transform the time
dependent integrals of motion into time independent integrals of motion and
integrate the latter and derive the general solution. In Sec.~\ref{sec4}, we repeat the
procedure for the case $q=2$ in Eq.~(\ref{int08}) and identify the integrable
systems. In Sec. \ref{sec5}, we consider the case $q=$ arbitrary in (\ref{int08}) and
unearth several new integrable equations and their associated mathematical
structures. Finally, we present our conclusions in Sec. \ref{sec6}.

\section{Generalized Extended Prelle-Singer (PS) procedure}
\label{sec2}
In this section we briefly recall the generalized extended or modified PS 
procedure before applying it to the specific problem in hand.
Sometime ago, Prelle and Singer$\footnotesize^5$ have proposed a 
procedure for solving first order ODEs that admit solutions in terms 
of elementary functions if such solutions exist.  The attractiveness 
of the PS method is that if the given system of first order ODEs has a solution 
in terms of elementary functions then the method guarantees that this solution 
will be found.  Very recently Duarte et al.$\footnotesize^{7,8}$ have modified 
the technique developed by Prelle and Singer$\footnotesize^{5,6}$ and applied 
it to second order ODEs.   
Their approach was based on the conjecture that if an elementary solution 
exists for the given second order ODE then there exists at least one 
elementary first integral $I(t,x,\dot{x})$ whose derivatives are all 
rational functions of $t$, $x$ and $\dot{x}$.  For a class of  
systems these authors have deduced first integrals and in some cases for the 
first time through their procedure$\footnotesize^7$. Recently the present 
authors have generalized the theory of Duarte et al.$\footnotesize^7$ in 
different directions and shown that for the second order ODEs one can isolate 
even two independent integrals of motion$\footnotesize^{9-11}$
and obtain general solutions explicitly without any integration. This theory 
has also been illustrated for a class of problems$\footnotesize^{1,9-11}$.
The authors have also generalized the theory successfully to higher order 
ODEs$\footnotesize^{10,17}$. For
example, in the case of third order ODEs the theory has been appropriately
generalized to yield three independent integrals of motion unambiguously so
that the general solution follows immediately from these integrals of 
motion$\footnotesize^{17}$. 

 We stress that the PS procedure has many advantages over other methods. To name
 a few, we cite: 
(1) For a given problem if the solution exists it has been conjectured 
that the PS method guarantees to provide first integrals.  (2) The PS method 
not only gives the first 
integrals but also the underlying integrating factors, that is, multiplying 
the equation with these functions we can rewrite the equation as a  perfect 
differentiable function which upon integration gives the first integrals directly. 
(3) The PS method can be used 
to solve nonlinear as well as linear second order ODEs. (4) As the PS method 
is based on the equations of motion rather than on Lagrangian or Hamiltonian
description,
the analysis is applicable to deal with both Hamiltonian and non-Hamiltonian 
systems. 

\subsection{PS method}
\label{sec21}
Let us rewrite Eq.~(\ref{int08}) in the form 
\begin{eqnarray} 
\ddot{x}=-((k_1x^q+k_2)\dot{x}+k_3x^{2q+1}+k_4x^{q+1}+\lambda_1x)
\equiv \phi(x,\dot{x}).
 \label{met1}
\end{eqnarray}
Further, we assume that the ODE (\ref{met1}) 
admits a first integral $I(t,x,\dot{x})=C,$ with $C$ constant on the 
solutions, so that the total differential becomes
\begin{eqnarray}  
dI={I_t}{dt}+{I_{x}}{dx}+{I_{\dot{x}}{d\dot{x}}}=0, 
\label{met3}  
\end{eqnarray}
where each subscript denotes partial differentiation with respect 
to that variable. Rewriting Eq.~(\ref{met1}) in the form 
$\phi dt-d\dot{x}=0$ and adding a null term 
$S(t,x,\dot{x})\dot{x}$ $ dt - S(t,x,\dot{x})dx$ to the latter, we obtain that on 
the solutions the 1-form
\begin{eqnarray}
\bigg(\phi +S\dot{x}\bigg) dt-Sdx-d\dot{x} = 0. 
\label{met6} 
\end{eqnarray}	
Hence, on the solutions, the 1-forms (\ref{met3}) and 
(\ref{met6}) must be proportional.  Multiplying (\ref{met6}) by the 
factor $ R(t,x,\dot{x})$ which acts as the integrating factors
for (\ref{met6}), we have on the solutions that 
\begin{eqnarray} 
dI=R(\phi+S\dot{x})dt-RSdx-Rd\dot{x}=0. 
\label{met7}
\end{eqnarray}
Comparing Eq.~(\ref{met3}) 
with (\ref{met7}) we have, on the solutions, the relations 
\begin{eqnarray} 
 I_{t}  = R(\phi+\dot{x}S),\quad 
 I_{x}  = -RS, \quad 
 I_{\dot{x}}  = -R.  
 \label{met8}
\end{eqnarray}
Then the compatibility conditions, 
$I_{tx}=I_{xt}$, $I_{t\dot{x}}=I_{{\dot{x}}t}$, $I_{x{\dot{x}}}=I_{{\dot{x}}x}$, 
between the Eqs.~(\ref{met8}), provide us 
\begin{eqnarray}          
S_t+\dot{x}S_x+\phi S_{\dot{x}} &=& 
   -\phi_x+\phi_{\dot{x}}S+S^2,\label {lin02}\\
R_t+\dot{x}R_x+\phi R_{\dot{x}} & =&
-(\phi_{\dot{x}}+S)R,\label {lin03}\\
R_x-SR_{\dot{x}}-RS_{\dot{x}}  &= &0.
\qquad \qquad\qquad \;\;\;\label {lin04}
\end{eqnarray}

Solving Eqs.~(\ref{lin02})-(\ref{lin04}) one can obtain expressions for $S$ and
$R$. It may be noted that any set of special solutions $(S,R)$ is sufficient for
our purpose. Once these forms are determined the integral of motion 
$I(t,x,\dot{x})$ can be deduced from the relation 
\begin{eqnarray}
 I= r_1
  -r_2 -\int \left[R+\frac{d}{d\dot{x}} \left(r_1-r_2\right)\right]d\dot{x},
  \label{met13}
\end{eqnarray}
where 
\begin{eqnarray} 
r_1 = \int R(\phi+\dot{x}S)dt,\quad
r_2 =\int (RS+\frac{d}{dx}r_1) dx. \nonumber
\end{eqnarray}
Equation~(\ref{met13}) can be derived straightforwardly by 
integrating the Eq.~(\ref{met8}).

The crux of the problem lies in finding the explicit solutions satisfying all
the three determining Eqs.~(\ref{lin02})-(\ref{lin04}), since once a 
particular solution is known the integral of  motion can be readily 
constructed. The difficulties in constructing admissible set of solutions
$(S,R)$ satisfying all the three Eqs.~(\ref{lin02})-(\ref{lin04}) and possible
ways of obtaining the solutions have been discussed in detail in Ref. 9. 
 
\section{Application of PS procedure to Eq.~(\ref{int07})}
\label{sec3}
Let us first consider the case $q=1$ in Eq.~(\ref{int08}) or equivalently
(\ref{int07})
\begin{eqnarray}            
\qquad \qquad\qquad \ddot{x}+(k_1x+k_2)\dot{x}+k_3x^3+k_4x^2+\lambda_1x=0.\qquad 
\qquad \qquad
\qquad \qquad \qquad \qquad 
\nonumber 
(\ref {int07})
\end{eqnarray}
Eq.~(\ref{int07}) itself includes several physically important models. For
example, choosing $k_i=0,\;i=1,...4$, we get the simple harmonic oscillator
equation and the choice $k_1,\;k_2=0$ gives us the anharmonic oscillator equation. 
When $k_1,\;k_4=0$ Eq.~(\ref{int07}) becomes the force-free Duffing oscillator 
equation$\footnotesize^{12}$. The choice $k_2,\;k_4,\;\lambda_1=0$ provides us 
the MEE$\footnotesize^{18}$. In the limit 
$k_3=\frac{k_1^2}{9},\;k_4=\frac{k_1k_2}{3}$, 
Eq.~(\ref{int07}) becomes MEE with linear term which is another 
linearizable equation which we have studied extensively in Refs. 9 and 19. 
The restriction $k_1,\;k_3=0$ leads us to the 
force-free Helmhotz oscillator$\footnotesize^{12,13}$. In the following we 
investigate whether the system (\ref{int07}) admits any other integrable case
besides the above.

We solve Eq.~(\ref{int07}) through the extended PS procedure in the
following way. For a given second order equation, (\ref{int07}), 
the first integral $I$ should
be either a time independent or time dependent one. In the former case, it is a
conservative system and we have $I_t=0$ and in the later case we have $I_t\neq0$. 
So let us first consider the case $I_t=0$ and determine the null forms and the 
corresponding
integrating factors and from these we construct the integrals of motion 
and then we do extend the analysis for the case $I_t\neq0$.
\subsection{The case $I_t=0$}
\label{sec31}
\subsubsection{Null forms}
\label{sec311}
In this case one can easily fix the null form $S$ from the first equation in
(\ref{met8}) as 
\begin{equation}
S = \frac{-\phi}{\dot{x}}=
-\frac{((k_1x+k_2)\dot{x}+k_3x^3+k_4x^2+\lambda_1x)}{\dot{x}}.
\label{mlin06}
\end{equation}
\subsubsection{Integrating Factors}
\label{sec312}
Substituting this form of $S$, given in (\ref{mlin06}), into (\ref{lin03}) we get 
\begin{eqnarray}          
 &R_t+\dot{x}R_x-((k_1x+k_2)\dot{x}+k_3x^3+k_4x^2+\lambda_1x)R_{\dot{x}} 
\qquad\qquad\nonumber\\&\qquad\qquad\qquad\qquad
=\bigg((k_1x+k_2)+\frac{((k_1x+k_2)\dot{x}+k_3x^3+k_4x^2+\lambda_1x)}
{\dot{x}}\bigg)R.\label {lin03a}
\end{eqnarray}

Equation~(\ref{lin03a}) is a first order linear partial differential equation
with variable coefficients. As we noted earlier any particular solution is
sufficient to construct an integral of motion (along with the function $S$).
To seek
a particular solution for $R$ one can make a suitable ansatz instead of looking
for the general solution. We assume $R$ to be of the form,
\begin{equation}
R = \frac{\dot{x}}{(A(x)+B(x)\dot{x})^r},
\label{mlin08}
\end{equation} 
where $A$ and $B$ are functions of their arguments, and $r$ is a constant
which are all to be
determined. We demand the above form of  ansatz, (\ref{mlin08}), due to the following
reason. To deduce the first integral $I$ we assume a rational form for $I$, 
that is,
$I=\frac{f(x,\dot{x})}{g(x,\dot{x})}$, where $f$ and $g$
are arbitrary functions of $x$ and $\dot{x}$ and are independent of $t$. Since we
already assumed that $I$ is independent of $t$, we have,
$I_x=\frac{f_{x}g-fg_x}{g^2}$ and $I_{\dot{x}}=\frac{f_{\dot{x}}g-fg_{\dot{x}}}{g^2}$.
From (\ref{met8}) one can see that $R=I_{\dot{x}}=\frac{f_{\dot{x}}g-fg_{\dot{x}}}{g^2},
\;S=\frac{I_x}{I_{\dot{x}}}=\frac{f_xg-fg_x}{f_{\dot{x}}g-fg_{\dot{x}}}$ and
$RS=I_x$, so that the denominator of the function $S$ should be the numerator of the
function $R$. Since the denominater of $S$ is $\dot{x}$ (vide Eq.~(\ref{mlin06}))
we fixed the numerator of $R$ as $\dot{x}$. To seek a suitable function in the 
denominator initially one can consider an
 arbitrary form $R=\frac{\dot{x}}{h(x,\dot{x})}$. However, it is difficult to
 proceed with this choice of $h$. So let us assume that $h(x,\dot{x})$ is a 
 function which is polynomial in
 $\dot{x}$. To begin with let us consider the case where $h$ is linear in $\dot{x}$,
 that is, $h=A(x)+B(x)\dot{x}$. Since $R$ is in rational form while taking
 differentiation or integration the form of the denominator 
 remains same but the  power of the denominator decreases or 
 increases by a unit order from
 that of the initial one. So instead of considering $h$ to be of the form
 $h=A(x)+B(x)\dot{x}$, one may consider a more general form
 $h=(A(x)+B(x)\dot{x})^r$, where $r$ is a constant to be determined. 
 Such a generalized
 form of $h$ and so $R$ leads to several new integrable cases as we see below.

Substituting (\ref{mlin08}) into (\ref{lin03a}) and solving the resultant
equations, we arrive at the relation
\begin{equation}
r(\dot{x}(A_x+B_x\dot{x})+\phi B)=(A+B\dot{x})\phi_{\dot{x}}.
\label{mlin09}
\end{equation} 
Solving Eq.~(\ref{mlin09}) with $\phi=-
((k_1x+k_2)\dot{x}+k_3x^3+k_4x^2+\lambda_1x)$, we find nontrivial forms for the
functions $A$ and $B$ for two choices, namely, $(i)\;k_1,k_2$ arbitrary and 
$(ii)\;k_1=\mbox{arbitrary}, \;k_2=0$ with restrictions on other parameters as given below. The
respective forms of the functions and the restriction on the parameters are
\begin{subequations}
\begin{eqnarray}
 &&(i)\;\underline{ k_1,\;k_2: \mbox{arbitrary}} \qquad \qquad\qquad  \nonumber\\
 &&\qquad\qquad  A(x)=\frac{(r-1)b_0}{r}(\frac{k_1}{2}x^2+k_2x),\;\; B(x)=b_0=\mbox{constant},
 \;\;r=\mbox{constant},
 \nonumber\\ 
&&\qquad\qquad  k_3=\frac{b_0(r-1)}{2r^2}k_1^2, \quad 
k_4=\frac{3b_0(r-1)}{2r^2}k_1k_2,\quad 
\lambda_1=\frac{b_0(r-1)k_2^2}{r^2},
\label{mlin10}\\
&& (ii)\; \underline{k_1=\mbox{arbitrary}, \;k_2=0}\qquad\qquad \qquad  \nonumber\\  
&&\qquad\qquad  A(x)=\frac{(r-1)b_0}{2r}k_1x^2+\frac{r\lambda_1}{k_1},\quad B(x)=b_0,
\nonumber\\
&&\qquad\qquad   k_3=\frac{b_0(r-1)}{2r^2}k_1^2,\quad k_4=0,\quad 
 \lambda_1=\mbox{arbitrary parameter (here)}.
\label{mlin10a}
\end{eqnarray}
\end{subequations} 
We note that the case $(ii)$
cannot be derived from case $(i)$ by taking $k_2=0$. For example, choosing 
$k_2=0$ in (\ref{mlin10}) we get not only $k_4=0$ but also $\lambda_1=0$ 
whereas in the case
$(ii)$ we have the freedom $\lambda_1=$ arbitrary, so the cases (\ref{mlin10}) 
and (\ref{mlin10a}) are to be treated as separate. Making use of the forms of 
$A$ and $B$ from Eqs.~ (\ref{mlin10}) and
(\ref{mlin10a}) into (\ref{mlin08}), the integrating factor, `$R$', for the 
two cases can be obtained as
\begin{subequations}
\begin{eqnarray}
&& (i)\;\underline{k_1,\;k_2:\mbox{arbitrary}}
\qquad \qquad \qquad \qquad \qquad \nonumber\\
&&\qquad \qquad\qquad \qquad  R=\frac{\dot{x}}{\bigg[\frac{(r-1)}{r}(\frac{k_1}{2}x^2+k_2x)+\dot{x}\bigg]^r},
\;\;\;r\neq0 \label{mlin11a}\\
&& (ii)\;\;\underline{k_1=\mbox{arbitrary}, \;k_2=0}
\qquad \qquad \qquad \qquad \qquad \nonumber\\
&&\qquad \qquad\qquad \qquad  R=\frac{\dot{x}}{\bigg[\frac{(r-1)}{2r}k_1x^2
+\frac{r\lambda_1}{k_1}+\dot{x}\bigg]^r},\;\;\;r\neq0. \label{mlin11b}
\end{eqnarray}
We note that $b_0$ is a common parameter in the above and it is absorbed in
the definition of `$R$', see Eqs.~(\ref{lin03}) and (\ref{lin04}).
While deriving the above forms of $R$ (Eqs.~(\ref{mlin11a}) and 
(\ref{mlin11b})) we assumed that $r\neq 0$ and for the choice $r=0$ we obtain
consistent solution only if
both the parameters $k_1$ and $k_2$ become zero. Of course,
this sub-case can be treated as a trivial one since when $k_1, k_2=0$
the damping term in Eq.~(\ref{int07}) vanishes and the system becomes an
integrable anharmonic oscillator. In this trivial case we have the integrating 
factor of the form:
\begin{eqnarray}
&&  (iii)\;\underline{k_1,\;k_2=0}
\qquad \qquad \qquad \qquad \qquad \nonumber\\
&& \qquad \qquad \qquad \qquad R=\dot{x},\;\;\; r=0.
\label{mlin11c}
\end{eqnarray}
\label{mlin11}
\end{subequations}

Finally one has to check the compatibility of forms $S$ and $R$ with the third
Eq.~(\ref{lin04}). We indeed verified that the sets
\begin{subequations} 
\begin{eqnarray}
(i)\;&&S=-\frac{((k_1x+k_2)\dot{x}+\frac{(r-1)}{r^2}(\frac{k_1^3}{2}x^2
+\frac{3k_1k_2}{2}x^2+k_2^2x))}{\dot{x}},\nonumber\\
&&R=\frac{\dot{x}}{(\frac{(r-1)}{r}(\frac{k_1}{2}x^2+k_2x)+\dot{x})^r},\quad
k_1,k_2= \mbox{arbitrary},\;\;r\neq0\label {rcha1}\\
(ii)\;&&S=-\frac{(k_1x\dot{x}+\frac{(r-1)}{2r^2}k_1^2x^3
+\lambda_1x)}{\dot{x}},\nonumber\\
&&R=\frac{\dot{x}}{(\frac{(r-1)}{2r}k_1x^2
+\frac{r\lambda_1}{k_1}+\dot{x})^r},\quad k_1=\mbox{arbitrary},\;k_2=0,
\;\;r\neq0\label {rcha2}\\
  \mbox{ and}\qquad\qquad\nonumber\\
(iii)\;&&S=-\frac{(k_3x^3+k_4x^2+\lambda_1x)}{\dot{x}},\;\;\;
R=\dot{x}, \quad k_1,\;k_2=0,\label {rcha3} 
\end{eqnarray}
\end{subequations}
satisfy the Eq.~(\ref{lin04}) individually. As a consequence all the three pairs 
form compatible sets of solution for the Eqs.~(\ref{lin02})-(\ref{lin04}).

\subsubsection{Integrals of motion}
\label{sec313}
Having determined the explicit forms of $S$ and $R$
one can proceed to construct integrals of motion using the expressions 
(\ref{met13}).  
The parametric restrictions (\ref{mlin10}) and (\ref{mlin10a}) fix the equation
of motion (\ref{int07}) to the following specific forms,
\begin{subequations}
\begin{eqnarray}            
 (i)\;  &&\ddot{x}+(k_1x+k_2)\dot{x}+\frac{(r-1)}{2r^2}
\bigg(k_1^2x^3+3k_1k_2x^2+2k_2^2 x\bigg)=0,\;\;r\neq0,
\label {cha2}\\
 (ii)\; && \ddot{x}+k_1x\dot{x}+\frac{(r-1)k_1^2}{2r^2}x^3
+\lambda_1x=0,\;\;r\neq0,
\label {cha3}\\
(iii)\; && \ddot{x}+k_3x^3+k_4x^2+\lambda_1x=0,\; \;r=0.
\label {cha4}
\end{eqnarray}
\end{subequations}
In the above $k_1,\;k_2,\;k_3,\;k_4,\;\lambda_1$ and $r$ are arbitrary 
parameters.

We note that the transformation $x=y-\frac{k_2}{k_1}$ transforms equation 
(\ref{cha2}) to the form
\begin{eqnarray}            
\ddot{y}+k_1y\dot{y}+\frac{(r-1)k_1^2}{2r^2}y^3
-\frac{(r-1)k_2^2}{2r^2}y=0,&\;\;r\neq0.
\label {cha6}
\end{eqnarray}
Eq. (\ref{cha6}) is obtained from Eq.~(\ref{cha3}) by fixing 
$\lambda_1=-\frac{(r-1)k_2^2}{2r^2}$. So, hereafter, we consider 
Eq.~(\ref{cha2}) as a special case of Eq.~(\ref{cha3}) and so discuss only 
Eq.~(\ref{cha3}) as the general one. It may be noted that Eq.~(\ref{cha3}) 
includes several  known 
integrable cases. For example, the choice $r=3$ and $\lambda_1=0$ 
in Eq.~(\ref{cha3}) yields the MEE$\footnotesize^{18}$. On the other hand the 
choice $r=-1$ leads
us to the equation $\ddot{x}+k_1x\dot{x}-k_1^2x^3+\lambda_1x=0$ which can be 
solved
in terms of Weierstrass elliptic function$\footnotesize^{20}$. {\it The other 
choices of $r$ lead to new integrable cases} as we see below.

Substituting the forms of $S$ and $R$ 
(vide Eqs.~(\ref{rcha2}) and (\ref{rcha3})) into the 
general 
form of the integral of motion (\ref{met13}) and evaluating the resultant 
integrals, we obtain the following time independent first integrals for the 
cases (\ref{cha3}) and (\ref{cha4}):
\begin{subequations}
\begin{eqnarray}
 (iia) \quad &&I_1=
\bigg(\dot{x}+\frac{(r-1)}{2r}k_1x^2+\frac{r\lambda_1}{k_1}
\bigg)^{-r} \label {intm03}\\  &&\qquad
\times \bigg[\dot{x}(\dot{x}+\frac{k_1}{2}x^2+\frac{r^2\lambda_1}{(r-1)k_1})
+\frac{(r-1)}{r^2}(\frac{k_1}{2}x^2+\frac{r^2\lambda_1}{(r-1)k_1})^2
\bigg],
\;\; r\neq0,1,2,\nonumber\\
 (iib) \quad 
&&I_1=\frac{4k_1\dot{x}}{k_1^2x^2+4k_1\dot{x}+8\lambda_1}
-\log(k_1^2x^2+4k_1\dot{x}+8\lambda_1),\;\; r=2,\label {intm04}\\ 
 (iic) \quad 
&&I_1=\dot{x}+\frac{k_1}{2}x^2
-\frac{\lambda_1}{k_1}\log(k_1\dot{x}+\lambda_1),\;\; r=1,\label {intm05}\\   
 (iii) \quad 
&&I_1=\frac{\dot{x}^2}{2}+\frac{k_3}{4}x^4+\frac{k_4}{3}x^3
+\frac{\lambda_1}{2}x^2, \;\; r=0.\label {intm06}
\end{eqnarray}
\label{mlin12}
\end{subequations}
Note that in Eq.~(\ref{intm03}), $r$ can take any real value, except $0,1,2$.
In the above integrals $I_1$ given by Eqs.~(\ref{intm03}) - (\ref{intm05}) 
correspond to the ODE (\ref{cha3}), while (\ref{intm06}) corresponds to 
the Eq.~(\ref{cha4}).

Due to the fact that the integrals of motion (\ref{mlin12}) are time independent, one can
look for a Hamiltonian description for the respective equations of motion. In fact, we
obtain the explicit Hamiltonian forms for all the above cases.
\subsubsection{Hamiltonian Description of (\ref{mlin12})}
\label{sec314}
Assuming the existence of a Hamiltonian
\begin{eqnarray}
I(x,\dot{x}) = H(x,p) = p \dot{x}-L(x,\dot{x}),
\label{mlin13}
\end{eqnarray}
where $L(x,\dot{x})$ is the Lagrangian and $p$ is the canonically conjugate
momentum, we have
\begin{eqnarray}
&&\frac{\partial{I}}{\partial\dot{x}} = \frac{\partial{H}}{\partial\dot{x}} = 
\frac{\partial{p}}{\partial\dot{x}}\dot{x} + p - \frac{\partial{L}}{\partial\dot{x}} = 
\frac{\partial{p}}{\partial\dot{x}} \dot{x}, \nonumber \\
&&\frac{\partial{I}}{\partial{x}} = \frac{\partial{H}}{\partial{x}} = 
\frac{\partial{p}}{\partial{x}}\dot{x} - \frac{\partial{L}}{\partial{x}}.
\label{mlin13a}
\end{eqnarray}
From (\ref{mlin13a}) we identify
\begin{eqnarray}
&&p =\int \frac{I_{\dot{x}}}{\dot{x}} d\dot{x},\nonumber \\ 
&&L = \int (p_x\dot{x}-I_{x})dx 
+ \int [p-\frac{d}{d\dot{x}}\int (p_x\dot{x}-I_{x})dx]d\dot{x}.
\label{mlin13b}
\end{eqnarray}

Plugging the expressions (\ref{mlin13}) into (\ref{mlin13b}) one can evaluate
the canonically conjugate momentum and the associated Lagrangian as well as the
Hamiltonian. They read as
follows:
\begin{subequations}
\begin{eqnarray}
  (a)\;&&\underline{\mbox{The canonical momenta}}: \nonumber\\
 &&\qquad(iia,b) \quad p=\frac{1}{r-1}\bigg(\dot{x}+\frac{(r-1)}{r}\frac{k_1}{2}x^2
+\frac{r\lambda_1}{k_1}\bigg)^{1-r}, \;\; r\neq0,1\;\;\qquad\qquad\\
\label{mlin12h}
 &&\qquad(iic) \quad  
p=\log(k_1\dot{x}+\lambda_1),\;\;r=1\label{mlin12j}\\ 
 &&\qquad(iii) \quad 
p=\dot{x}, \;\; r=0.\label{mlin12k}
\end{eqnarray}
\end{subequations}
(Note in the above $r=2$ is included in Eq.~(\ref{mlin12h}) itself).
\begin{subequations}
\begin{eqnarray}
  (b)\;&&\underline{\mbox{The Lagrangian}}: \nonumber\\
 &&\qquad(iia) \quad L=\frac{1}{(2-r)(r-1)}\bigg(\dot{x}+\frac{(r-1)}{r}\frac{k_1x^2}{2}
+\frac{r\lambda_1}{k_1}\bigg)^{2-r}, \;\; r\neq0,1,2
\label{lag01}\\
&& \qquad (iib) \quad  
L=\log(4k_1\dot{x}+8\lambda_1+k_1^2x^2), \;\; r=2\label{lag01a}\\
 &&\qquad(iic) \quad  
L=\frac{\lambda_1}{k_1}\log(k_1\dot{x}+\lambda_1)
+\dot{x}(\log(k_1\dot{x}+\lambda_1)-1)-\frac{1}{2}k_1x^2,\;\;r=1\label{lag02}\\ 
 &&\qquad(iii) \quad 
L=\frac{\dot{x}^2}{2}-\frac{k_3}{4}x^4-\frac{k_4}{3}x^3-\frac{\lambda_1}{2}x^2, 
\;\; r=0.\label{lag03}
\end{eqnarray}
\label{com02a}
\end{subequations} 
\begin{subequations}
\begin{eqnarray}
  (c)\;&&\underline{\mbox{The Hamiltonian}}: \nonumber\\
 &&\qquad(iia) \quad  
H=\bigg[\frac{\bigg((r-1)p\bigg)^{\frac{r-2}{r-1}}}{(r-2)}-p(
\frac{(r-1)}{2r}k_1x^2+\frac{r\lambda_1}{k_1})\bigg],
\;\; r\neq0,1,2\label{mlin12d}\\
 &&\qquad(iib) \quad 
H=\frac{2\lambda_1}{k_1}p+\frac{k_1}{4}x^2p+
\log(\frac{4k_1}{p}),\;\; r=2\label{mlin12e}\\
 &&\qquad(iic) \quad  
H=\frac{1}{k_1}(e^{p}-\lambda_1p+\frac{k_1^2}{2}x^2-\lambda_1),
\;\;r=1\label{mlin12f}\\ 
 &&\qquad(iii) \quad 
H=\frac{p^2}{2}+\frac{k_3}{4}x^4+\frac{k_4}{3}x^3+\frac{\lambda_1}{2}x^2,
 \;\; r=0.\label{mlin12g}
\end{eqnarray}
\label{mlin012}
\end{subequations}
One can check that the Hamilton's equations of motion are indeed equivalent 
to the appropriate equation (\ref{int07}).

Since Eqs.~(\ref{cha3}) and (\ref{cha4}) admit time independent 
Hamiltonians they can be classified as Liouville integrable systems. 
{\it The important fact we want to stress here is that for arbitrary values 
of $r$, including fractional values, the equation
(\ref{cha3}) is integrable}.

\subsubsection{Canonical transformation for the Hamiltonian 
Eqs.~(\ref{mlin012})}
\label{sec315}
Interestingly, we also identifed suitable canonical transformation to standard
particle in a potential description for the 
Hamiltonians (\ref{mlin012}). Now introducing the canonical 
transformations 
\begin{eqnarray}
x=&\frac{2rP}{k_1U},\qquad\;\;
p=-\frac{k_1U^2}{4r},&\;\; r\neq0,1,
\label {com05}\\
x=&\frac{P}{k_1},\qquad\;\; p=-k_1U,&\;\; r=1
\label {com05a}
\end{eqnarray}
the Hamiltonian $H$ in Eq.~(\ref{mlin012}) 
can be recast in the standard form (after rescaling)
\begin{eqnarray}
H=\left\{
\begin{array}{ll}
\frac{1}{2}P^2
+\frac{(1-r)}{(r-2)}\bigg(\frac{(r-1)k_1U^2}{4r}\bigg)^{\frac{(r-2)}{r-1}}
+\frac{(r-1)\lambda_1}{4}U^2,&\;r\neq0,1,2\\\\
\frac{1}{2}P^2+\frac{\lambda_1}{4}U^2+
\log(\frac{32}{U^2}),&\;r=2\\\\
\frac{1}{2}P^2+e^{-k_1U}+\lambda_1k_1U,&\;r=1\\\\
\frac{1}{2}P^2+\frac{k_3}{4}U^4+\frac{\lambda_1}{2}U^2,&\; r=0.
\end{array}\right.\label{com06}
\end{eqnarray} 
It is straightforward to check that when $U$ and $P$ are canonical so do $x$ 
and $p$ (and vice versa) and the corresponding equations of motion 
turn out to be 
\begin{subequations} 
\begin{eqnarray}
&& \ddot{U}-2\bigg(\frac{(r-1)k_1}{4r}\bigg)^{\frac{(2-r)}{(1-r)}}
U^{\frac{(3-r)}{(1-r)}}+\frac{(r-1)\lambda_1}{2}U=0, \;\; r\neq0,1
\label{eq01}\\
&& \ddot{U}-k_1e^{-U}+k_1\lambda_1=0,
\qquad\qquad\qquad\qquad\qquad \quad \;\;r=1\label{eq03}\\ 
&&  \ddot{U}+k_3U^3+\lambda_1U=0,
\qquad \qquad\qquad\qquad\qquad\quad\;\;\;\; r=0.\label{eq04}
\end{eqnarray}
\label{com07}
\end{subequations}
One may note that the equations of motion now become standard type 
anharmonic oscillator equations.

\subsection{The case $I_t\neq0$}
\label{sec32}
In the previous sub-section we considered the case $I_t=0$. As a consequence $S$
turns out to be $\frac{-\phi}{\dot{x}}$. However in the case $I_t=0$, the function $S$
has to be determined from Eq.~(\ref{lin02}), that is,
\begin{eqnarray}
&&S_t+\dot{x}S_x-((k_1x+k_2)\dot{x}+k_3x^3+k_4x^2+\lambda_1x) S_{\dot{x}} 
\nonumber\\
  &&\qquad\qquad \qquad=(k_1\dot{x}+3k_3x^2+2k_4x+\lambda_1)-(k_1x+k_2)S+S^2.
   \label{mlin05a}
\end{eqnarray}
Since it is too difficult to solve Eq.~(\ref{mlin05a}) for its general solution, 
we seek a
particular solution for $S$, which is sufficient for our purpose. In
particular, we seek a simple rational expression for $S$ in
the form 
\begin{eqnarray}
S = \frac{a(t,x)+b(t,x)\dot{x}}{c(t,x)+d(t,x)\dot{x}},
\label{mlin05}
\end{eqnarray}
where $a,\;b,\;c$ and $d$ are arbitrary functions of $t$ and $x$ which are to be
determined. Of course, the analysis of this form alone does not exhaust all
possible cases of interest. We hope to make a more exhaustive study of
Eq.~(\ref{mlin05a}) separately.
Substituting (\ref{mlin05}) into (\ref{mlin05a}) and equating the coefficients 
of different powers of $\dot{x}$ to zero, we get
\begin{eqnarray}          
 && db_x-bd_x -k_1d^2  = 0 ,\nonumber\\
 && db_t-bd_t+cb_x-bc_x+a_xd-ad_x 
-2k_1cd-(3k_3x^2+2k_4x+\lambda_1)d^2
\nonumber\\&&\qquad\qquad\qquad\qquad\qquad\qquad\qquad\qquad
+(k_1x+k_2)bd-b^2  = 0,\nonumber\\
 && cb_t-bc_t+da_t-ad_t+ca_x-ac_x 
-k_1c^2-2(3k_3x^2+2k_4x+\lambda_1)cd
\nonumber\\&&\qquad\qquad\qquad\qquad\qquad\qquad\qquad\qquad
+2(k_1x+k_2)ad-2ab  = 0,\nonumber\\
 && ca_t-ac_t-(k_3x^3+k_4x^2+\lambda_1x)(bc-ad) 
-(3k_3x^2+2k_4x+\lambda_1)c^2
\nonumber\\&&\qquad\qquad\qquad\qquad\qquad\qquad\qquad\qquad
+(k_1x+k_2)ac-a^2 = 0.
\label{lam106}
\end{eqnarray}
The determining equation for the functions $a,b,c$ and $d$ have now turned out
to be nonlinear. To solve these equations we further assume that the functions
$a,b,c$ and $d$ are polynomials in $x$ with coefficients which are arbitrary
functions in $t$. Substituting these forms into Eqs.~(\ref{lam106}) we
obtain another enlarged set of determining equations for the unknowns 
and solving the latter consistently 
we obtain nontrivial solutions for the functions $a,b,c$ and $d$ for four 
sets of parametric 
choices. We present the explicit forms of the associated null function $S$ 
given by (\ref{mlin05}) and the
parametric restrictions in Table I.

Now substituting the forms of $S$ into Eq.~(\ref{lin03}) and solving the
resultant equation we obtain the corresponding forms of $R$. To solve the
determining equation for $R$ we again seek the same form of ansatz (\ref{mlin08}) but
with explicit $t$ dependence on the coefficient functions, that is, 
$R = \frac{S_d}{(A(t,x)+B(t,x)\dot{x})^r},$ 
where $S_d$ is the denominator of $S$. We report the
resultant forms of $R$ in Table I. Once $S$ and $R$
are determined then one has to verify the compatibility of this set $(S,R)$ 
with the extra constraint Eq.~(\ref{lin04}). We find that the forms $S$ and $R$ 
given in Table I do satisfy the
extra constraint equation and form a compatible solution. Now substituting
$S_i$'s and $R_i$'s into Eq.~(\ref{met13}) one can construct the associated
integrals of motion. We report the integrals of motion $(I)$ in Table I along
with the forms $S$ and $R$.

\begin{sidewaystable}
\caption{ Parametric restrictions, null forms $(S)$, integrating factors $(R)$ 
and time dependent integrals of motion $(I)$ of\\
\centerline{$\ddot{x}+(k_1x+k_2)\dot{x}+k_3x^3+k_4x^2+\lambda_1x=0$ (identified
with the assumed ansatz form of $S$ and $R$)}}
\small
\begin{tabular}{|p{.9cm}|p{4.2cm}|p{3.3cm}|p{3.9cm}|p{7cm}|}
\hline
Cases& Parametric restrictions & Null form $(S)$ & Integrating factor $(R)$ & 
Integrals of motion $(I)$  \\
\hline
(i) &$k_3=\frac{k_1^2}{9},\;k_4=\frac{k_1k_2}{3}$ & & &
$(a)\;I=e^{\mp\omega t}
\left(\frac{3\dot{x}-\frac{3(-k_2\mp\omega)}{2}x+k_1x^2}
{3\dot{x}-\frac{3(-k_2\pm\omega)}{2}x+k_1x^2}\right),
$ \\
& 
&$ \frac{(\frac{k_1}{3}x^2-\dot{x})}{x}$
&$\displaystyle{\frac{xe^{\mp\omega t}}
{(\dot{x}-\frac{(k_2\pm\omega)}{2}x+\frac{k_1}{3}x^2)^2}}$
&$\;\;\;\;\;\;\;\;\;\;k_2,\lambda_1\neq0,\;\omega=(k_2^2-4\lambda_1)^{\frac{1}{2}}$\\
&($k_1,\;k_2,\;\lambda_1:\;$arbitrary)&&&
$(b)\;I=-t+\frac{x}{(\frac{k_2}{2}x+\frac{k_1}{3}x^2+\dot{x})},
\;\;\;\;\;\; k_2^2=4\lambda_1$\\
\hline
(ii) & $k_3=0,k_4=\frac{k_1}{4}(k_2\pm\omega),$&
$\displaystyle{\frac{1}{2}(k_2\mp\omega)+k_1x},$  &
$\displaystyle{e^{\frac{(k_2\pm\omega)}{2}t}}$  &
$I=\bigg(\dot{x}+\frac{k_2\mp\omega}{2}x
+\frac{k_1}{2}x^2\bigg)
e^{(\frac{k_2\pm\omega}{2})t},$  \\
&($k_1,\;k_2,\;\lambda_1:\;$arbitrary)
&&&$\;\;\;\;\;\;\;\;\;\;\omega=(k_2^2-4\lambda_1)^{\frac{1}{2}}$\\
\hline
(iii) & $k_1,\;k_3=0,\;\lambda_1=\frac{6k_2^2}{25}$ & 
$\displaystyle{\frac {(\frac{2k_2 \dot{x}}{5}+\frac{4k_2^2 x}{25}+{k_4 x^2})}
{ (\dot{x}+\frac {2k_2}{5}x)}}$ &
$\displaystyle{(\dot{x}+\frac {2k_2}{5}x)e^{\frac{6}{5}k_2 t}}$ &
$I=e^{\frac{6}{5}k_2 t} \bigg(\frac{\dot{x}^2}{2}
+\frac{2k_2}{5} x\dot{x}+\frac{2k_2^2}{25} x^2+\frac{k_4 }{3}x^3\bigg)$ \\
&($k_2,\;k_4:\;$arbitrary)&&&\\
\hline
(iva) & $k_3=\frac{(r-1)k_1^2}{2r^2},\;k_4=\frac{k_1k_2}{3},$ &
& 
&
$I=\bigg(\frac{k_3}{2}x^4
+(\dot{x}+\frac{k_2}{3}x)(\dot{x}+\frac{k_2}{3}x+\frac{k_1}{2}x^2)\bigg)$
\\
&$\lambda_1=\frac{2k_2^2}{9},\;r\neq0$&
$\frac{k_2}{3}+k_1x+\frac {3k_3x^3}{(3\dot{x}+k_2x)}$ &
$\displaystyle{\frac{(k_2x+3\dot{x})e^{(\frac{2(2-r)k_2}{3})t}}
{(\frac{k_2}{3}x+rk_3x^2+\dot{x})^{r}}}$&
$\;\;\;\times \bigg(\dot{x}+\frac{k_2}{3}x
+rk_3x^2\bigg)^{-r}e^{\frac{2(2-r)}{3}k_2 t},\;\;r\neq2$ 
\\
&($k_1,\;k_2,\;r:\;$arbitrary)
&&&$I=\frac{2}{3}k_2t+\log(4k_2x+3k_1x^2+12\dot{x})$
\\
&&&&$\;\;\;\;\displaystyle{-\frac{4(k_2x+3\dot{x})}{(4k_2x+3k_1x^2
+12\dot{x})}},
\;\;\;\;\;\;\;\;\;\;\;\;\;\;r=2$\\
\hline
(ivb) & $k_1=0,\;k_4=0,$ &
$\displaystyle{\frac{(\frac{k_2 }{3}\dot{x}+\frac{k_2^2 }{9}x+k_3x^3)}
{(\dot{x}+\frac{k_2 x}{3})}}$ &
$\displaystyle{e^{\frac{4}{3}k_2 t}{(\dot{x}+\frac{k_2 x}{3})}}$ &
$I=e^{\frac{4}{3}k_2 t} \bigg[\frac{\dot{x}^2}{2}
+\frac{k_2 }{3}x\dot{x}+\frac{k_2^2 }{18}x^2+\frac{k_3}{4}x^4\bigg]$ \\
&$\lambda_1=\frac{2k_2^2}{9},r=0$&&&\\
&($k_2,\;k_3:\;$arbitrary)
&&&
\\
\hline
\end{tabular}
\end{sidewaystable}
At this stage, we note that the first integral for the {\it case} $(i)$ with 
$k_2,\lambda_1=0$ has been derived 
in Ref. 18 through Lie symmetry analysis. However, recently, we have
derived$\footnotesize^{9}$ the first integral for arbitrary values of $k_2$ and
$\lambda_1$. {\it The case $(ii)$ is new to the literature}. The first integral for 
the case $(iii)$ was reported recently in Refs. 9,12 and 13. 
{\it The first integral for the case $(iva)$ is new to the 
literature}. The {\it case} $r=0$ discussed as $(ivb)$ is nothing but the 
force-free Duffing oscillator whose integrability has been discussed in 
Refs.~12 and 14. 
 
Since we obtained only one integral in each case, (except case $(i)$ where we
have found second explicit time dependent integral, see Ref. 9),
which are also time dependent ones, we
need to integrate them further to obtain the second integration constant and
prove the complete integrability of the respective systems, which is indeed a 
difficult task. 

In this connection we have introduced a new method$\footnotesize^{1,9}$ which can be 
effectively used to transform the
time dependent integral into a time independent one, for a {\it class of
problems}, so
that the latter can be integrated easily. We invoke this 
procedure here in order to integrate the time dependent first integrals and 
obtain the general
solution for all the cases in Table I (except case $(iv)$, see below). For the 
{\it case} $(iv)$, we prove the Liouville integrability of it.
\subsection{Method of transforming time dependent first integral to time
independent one}
\label{sec33}
Let us assume that there exists a first integral for the equation 
(\ref{int07}) of the form,
\begin{eqnarray}  
I=F_1(t,x,\dot{x})+F_2(t,x). \label{the01}
\end{eqnarray}
Now  let us split the function $F_1$ further in terms of two functions such 
that $F_1$ itself is a function of the product of the two functions, say, a 
perfect differentiable function $\frac {d}{dt}G_1(t,x)$ and another function 
$G_2(t,x,\dot{x})$, that is,
\begin{eqnarray} 
I=F_1\left(\frac{1}{G_2(t,x,\dot{x})}\frac{d}{dt}G_1(t,x)\right)
+F_2\left(G_1(t,x)\right),
\label{the02}
\end{eqnarray}
where $F_1$ is a function which involves the variables $t,x$ and $\dot{x}$
whereas $F_2$ should involve only the variable $t$ and $x$.
We note that while rewriting Eq.~(\ref{the01}) in the form 
(\ref{the02}), we demand that the function $F_2(t,x)$ in (\ref{the01}) 
automatically to be a function of $G_1(t,x)$.  Now identifying  the function  
$G_1$ as the new dependent variable and the integral of $G_2$ over time as 
the new independent variable, that is,
\begin{eqnarray} 
w = G_1(t,x),\quad z = \int_o^t G_2(t',x,\dot{x}) dt', 
\label{the03}
\end{eqnarray}
one indeed obtains an explicit transformation to remove  the time dependent 
part in the first integral.  We note here that the integration 
on the right hand side of (\ref{the03}) leading to $z$ can be performed 
provided the function $G_2$ is an exact derivative of $t$, that is,
$G_2=\frac{d}{dt}z(t,x)=\dot{x}z_x+z_t$, so that $z$ turns 
out to be a function
$t$ and $x$ alone.  In terms of the new variables, Eq.~(\ref{the02}) can 
be modified to the form
\begin{eqnarray} 
I=F_1\left(\frac {dw}{dz}\right)+F_2(w).
\label{the04}
\end{eqnarray}
In other words,
\begin{eqnarray} 
F_1\left(\frac {dw}{dz}\right)=I-F_2(w).
\label{the05}
\end{eqnarray}

Now rewriting Eq.~(\ref{the04}) one obtains a separable equation 
\begin{eqnarray}  
\frac {dw}{dz}=f(w),
\end{eqnarray}
which can lead to the solution after an integration.  Now rewriting the 
solution in terms 
of the original variables one obtains a general solution for the given equation.

In the following using the above idea we integrate the first integrals given in 
Table I and
deduce the second integration constant and general solution.
\subsection{Application}
\label{sec34}
$\bf{Case\;(ia)}$:$\;\underline{k_3=\frac{k_1^2}{9},\;k_4=\frac{k_1k_2}{3},
\;k_1,\;k_2\;\mbox{and}\;\lambda_1:\; \hbox{arbitrary}}$:\\

 The parametric restrictions given above fix the equation of motion
(\ref{int07}) in the form
\begin{eqnarray}
\ddot{x}+(k_1x+k_2)\dot{x}+\frac{k_1^2}{9}x^3+\frac{k_1k_2}{3}x^2+\lambda_1x=0,
\label {case01}
\end{eqnarray}
Let us rewrite the first integral associated for this case (vide case 
$(i)$ in Table I) in the form
\begin{eqnarray}
  \qquad I_1=-\frac{k_1e^{\frac{k_2\mp\omega}{2}t}x^2}
{(3\dot{x}-\frac{(-k_2\pm\omega)}{2}3x+k_1x^2)}\left[\frac {d}{dt}
\left((\frac{-3}{k_1x}+\frac{-k_2\pm\omega}
{2\lambda_1})e^{\frac{-k_2\mp\omega}{2}t}
\right)\right],
\label {the06} 
\end{eqnarray}
where $\omega=\sqrt{k_2^2-4\lambda_1}$. Comparing this with the equation
(\ref{the02}), and using (\ref{the03}), we obtain
\begin{eqnarray}
 w  = (\frac{-3}{k_1x}+\frac{-k_2\pm\omega}
{2\lambda_1})e^{\frac{-k_2\mp\alpha}{2}t},\quad 
z =(\frac{-3}{k_1x}+\frac{-k_2\mp\omega}
{2\lambda_1})e^{\frac{-k_2\pm\omega}{2}t}.
\label {the07}
\end{eqnarray}
Substituting (\ref{the07}) into Eq.~(\ref{case01}), the latter becomes the free
particle equation, namely, $\frac{d^2w}{dz^2}=0$, whose general solution is 
$w=I_1z+I_2$, where $I_1$ and $I_2$ are integration constants.  Rewriting 
$w$ and $z$ in terms of $x$ and $t$ one gets 
\begin{eqnarray} 
 x(t) =\left(\frac{6\lambda_1(1-I_1e^{\omega t})}
{k_1\omega(1+I_1e^{\omega t})
-(k_2\pm\omega)I_2e^{\frac{k_2\pm\omega}{2}t}
-k_1k_2(1-I_1e^{\omega t})}\right), 
\label{the07a}
\end{eqnarray}
where $\omega=\sqrt{k_2^2-4\lambda_1}$. 

Interestingly one can
consider several sub-cases. In the following we discuss some important ones
which are being  widely discussed in the current literature. In particular, the
difference in dynamics arises mainly depending on the sign of the parameter
$\alpha$ ($=\sqrt{k_2^2-4\lambda_1}$). We consider the cases 
$(i)$ $k_2^2< 4\lambda_1$ $(ii)$ $k_2^2>4\lambda_1$ and $(iii)$ 
$k_2^2=4\lambda_1$ separately. The restriction $k_2^2< 4\lambda_1$ 
reduces the solution (\ref{the07a}) to the form$\footnotesize^4$, 
\begin{eqnarray} 
x(t)=\frac{A\cos(\omega_0t+\delta)}{\bigg(e^{\frac{k_2}{2}t}
+\frac{2k_1A}{3(k_2^2+4\omega_0^2)}(2\omega_0\sin(\omega_0t+\delta)
-k_2\cos(\omega_0t+\delta))\bigg)},
\label{the07c}
\end{eqnarray}
where $\omega_0=\frac{\sqrt{4\lambda_1-k_2^2}}{2}$ and $\delta,\;A$ are 
arbitrary constants.
A further restriction $k_2=0$ gives us  the purely sinusoidally 
oscillating solution$\footnotesize^{19}$
\begin{eqnarray}
x(t)=\frac{A\sin{(\omega_0 t+\delta)}}
{1-(\frac{k}{3\omega_0})A\cos{(\omega_0 t+\delta)}},\quad
 0\leq A <\frac{3\omega_0}{k},
\quad \omega_0=\sqrt{\lambda_1},
\label{the07d}
\end{eqnarray}
where $A$ and $\delta$ are arbitrary constants. The associated equation of
motion, namely $\ddot{x}+k_1x\dot{x}+\frac{k_1^2}{9}x^3+\lambda_1x=0$, admits
very interesting nonlinear dynamics, see for example in Ref. 19.
 
On the other hand, in the limit $k_2^2> 4\lambda_1$ the solution looks like a 
dissipative/front-like one$\footnotesize^{19}$. A further restriction 
$\lambda_1=0$ takes us to the solution of the form$\footnotesize^{11}$
\begin{eqnarray} 
 x(t)=\bigg(\frac{3k_2(I_1e^{k_2t}-1)}
{k_1+k_2(3I_2+k_1I_1t)e^{k_2t}}\bigg).
\label{the07b}
\end{eqnarray}\\

$\bf{Case\;(ib)}$:$\;\underline{k_3=\frac{k_1^2}{9},\;k_4=\frac{k_1k_2}{3},
\;k_2^2=4\lambda_1,\; k_1\;\mbox{and}\;k_2:\; \mbox{arbitrary}}$:\\

The third choice $k_2^2= 4\lambda_1$ in (\ref{the07a}) leads us to the solution
\begin{eqnarray}
 x(t)=\bigg(\frac{3(I_1+t)}{3I_2e^{\frac{k_2}{2}t}-\frac{2k_1}{k_2^2}
 (2+I_1k_2+k_2t)}\bigg).
\label{the07f}
\end{eqnarray}
Further parametric restriction 
$k_2,\;\lambda_1=0$ provides us the general solution of the form
\begin{eqnarray}
 x(t)=\bigg(\frac{6(I_1+t)}{k_1(I_1+t)^2+6I_2}\bigg).
\label{the07e}
\end{eqnarray}
The underlying equation, that is, $\ddot{x}+k_1x\dot{x}+\frac{k_1^2}{9}x^3=0$,
is the $l=1$ integrable case of Eq.~(\ref{int05}) with the solution (\ref{hie05})
(see for example in Refs.~18 and 19). \\

$\bf{Case\;(ii)}$: $\;\underline{k_3=0,\;k_4=\frac{k_1}{4}(k_2\pm\sqrt{k_2^2
-4\lambda_1}),\;k_1,\;k_2\;\mbox{and}\;\lambda_1:\; \mbox{arbitrary}}$:\\ 

In this case we have the equation of the form 
\begin{eqnarray}
\ddot{x}+(k_1x+k_2)\dot{x}+\frac{k_1}{4}(k_2\pm\sqrt{k_2^2-4\lambda_1})x^2
+\lambda_1x=0.
\label {case02}
\end{eqnarray}
The associated first integral reads (vide case $(ii)$ in Table 1)
\begin{eqnarray} 
I=\bigg(\dot{x}+\frac{k_2\mp\sqrt{k_2^2-4\lambda_1}}{2}x
+\frac{k_1}{2}x^2\bigg)
e^{\frac{k_2\pm\sqrt{k_2^2-4\lambda_1}}{2}t}.
\label {the08}
\end{eqnarray}
Note that Eq.~(\ref{the08}) can be rewritten as a Riccati equation of the 
form$\footnotesize^{21}$
\begin{eqnarray} 
\dot{x}=Ie^{(\frac{-k_2\mp\sqrt{k_2^2-4\lambda_1}}{2})t}
-\bigg(\frac{k_2\mp\sqrt{k_2^2-4\lambda_1}}{2}\bigg)x-\frac{k_1}{2}x^2.
\label {rthe08}
\end{eqnarray}
The general solution of the Riccati equation is known to be free from movable
critical points and satisfies the Painlev\'e property. In this sense
Eq.~(\ref{case02}) can be considered as integrable in the Painlev\'e criteria
sense. However, in the general
case, (\ref{rthe08}), it is not clear whether it can be explicitly integrated
further. However, for the special case $\lambda_1=\frac{2k_2^2}{9}$ it can be
integrated as follows.

The restriction $\lambda_1=\frac{2k_2^2}{9}$ fixes the equation of motion
(\ref{case02}) and the first integral (\ref{the08}) in the forms
\begin{eqnarray}
\ddot{x}+(k_1x+k_2)\dot{x}+\frac{k_1k_2}{3}x^2+\frac{2k_2^2}{9}x=0,
\label {case02a}
\end{eqnarray}
and
\begin{eqnarray} 
I=\bigg(\dot{x}+\frac{k_2}{3}x+\frac{k_1}{2}x^2\bigg)
e^{\frac{2k_2}{3}t},
\label {the08a}
\end{eqnarray}
respectively. Now rewriting (\ref{the08a}) in the form (\ref{the02}), we get
\begin{eqnarray} 
I=e^{\frac{k_2}{3}t}\bigg(\frac {d}{dt}(xe^{\frac{k_2}{3}t})\bigg)
+\frac{k_1}{2}(xe^{\frac{k_2}{3}t})^2.
\label {the09}
\end{eqnarray}
Identifying the dependent and independent variables from (\ref{the09}) and
using the identities (\ref{the03}), we obtain the transformation
\begin{eqnarray} 
w = xe^{\frac{k_2}{3}t}, \quad
z = -\frac {3}{k_2}e^{-\frac{k_2}{3}t}. 
\label{the10}
\end{eqnarray}
Using the transformation (\ref{the10}) the first integral (\ref{the08a}) 
can be rewritten in the form
\begin{equation}
\hat{I} = w'+\frac{k_1}{2}w^2
\label{the11}
\end{equation}
which in turn leads to the solution by an integration, that is,
\begin{eqnarray} 
w(z) = \sqrt{\frac{2I}{k_1}}\tanh\bigg[{\sqrt{\frac{k_1I}{2}}(z-z_0)}\bigg],
\label{the011a}
\end{eqnarray}
where $z_0$ is arbitrary constant. Rewriting (\ref{the011a}) in terms of old variables we get
\begin{eqnarray} 
x(t) = \sqrt{\frac{2I}{k_1}}e^{-(\frac{k_2}{3})t}
\tanh\bigg[\frac {3}{k_2}(\sqrt{\frac{k_1I}{2}})(e^{-\frac{k_2}{3}t_0}
-e^{-\frac{k_2}{3}t})\bigg],
\label{the011b}
\end{eqnarray}
where $t_0$ is the second integration constant.\\

$\bf{Case\;(iii)}$: $\;\underline{k_1,\; k_3=0,\;\lambda_1=\frac{6k_2^2}{25},
\;k_2\;\mbox{and}\;k_4:\; \mbox{arbitrary}}$:\\

The corresponding equation of motion is 
\begin{eqnarray}
\ddot{x}+k_2\dot{x}+k_4x^2+\frac{6k_2^2}{25}x=0.
\label {case03}
\end{eqnarray}
Rewriting the associated first integral $I_1$, given in Case $(iii)$ in 
Table I, in the form 
(\ref{the01}), we get
\begin{eqnarray}
I=\frac{1}{2} \bigg(\dot{x}+\frac{2k_2 }{5}x\bigg)^2 e^{\frac{6}{5}k_2 t}
+ \frac{k_4 }{3}x^3e^{\frac{6}{5}k_2 t}.
\label{the15}
\end{eqnarray}
Now splitting the first term in Eq.~(\ref{the15}) further in the form 
(\ref{the02}), we obtain
\begin{eqnarray} 
I=e^{\frac{2k_2 t}{5}}\bigg(\frac {d}{dt}(\frac{1}{\sqrt{2}}x 
 e^{\frac {2k_2 t}{5}})\bigg)^2+\frac{k_4}{3}(xe^{\frac{2}{5}k_2 t})^3. 
 \label{the16}
\end{eqnarray}
Identifying the dependent and independent variables from (\ref{the16}) and 
using the relations (\ref{the03}), we obtain the transformation
\begin{eqnarray} 
w = \frac{1}{\sqrt{2}}x e^{\frac {2k_2 t}{5}}, \quad
z = -\frac {5}{k_2}e^{-\frac {k_2 t}{5}}. 
\label{the17}
\end{eqnarray}
Using this transformation, (\ref{the17}), the first integral (\ref{the15}) 
can be rewritten in the form
\begin{equation}
\hat{I} = w'^2+\frac{\hat{k_4}}{3}w^3,
\label{the18}
\end{equation}
where $\hat{k_4}=2\sqrt{2}k_4$, which inturn leads to  
\begin{eqnarray}
w'^2=4w^3-g_3,
\label{the19}
\end{eqnarray}
where $z=2\sqrt{\frac{3}{\hat{k_4}}}\hat{z}$
and $g_3=-\frac{12I_1}{\hat{k_4}}$. The solution of this differential 
equation can be represented in terms of Weierstrass
function$\footnotesize^{12,13}$ $\varrho(\hat{z};0,g_3)$.\\

$\bf{Case\;(iv)}$: $\;\underline{k_3=\frac{(r-1)}{2r^2}k_1^2,
\;k_4=\frac{k_1k_2}{3},\;\lambda_1=\frac{2k_2^2}{9},
\;k_1,\;k_2\;\mbox{and}\;r:\; \mbox{arbitrary\;(but not zero)}}$:\\

The above parameters fix the equation of motion (\ref{int07}) in the form
\begin{eqnarray}
\ddot{x}+(k_1x+k_2)\dot{x}+\frac{(r-1)k_1^2}{2r^2}x^3
+\frac{k_1k_2}{3}x^2+\frac{2k_2^2}{9}x=0,\;\; r\neq0.
\label {case04}
\end{eqnarray}
The associated first integral reads (vide case $(iva)$ in Table I)
\begin{eqnarray}
I=\left\{
\begin{array}{ll}
\bigg(\frac{(r-1)}{4r^2}k_1^2x^4
+(\dot{x}+\frac{k_2}{3}x)(\dot{x}+\frac{k_2}{3}x+\frac{k_1}{2}x^2)\bigg)
\bigg)\\\\
\qquad \qquad \times \bigg(\dot{x}+\frac{k_2}{3}x
+\frac{(r-1)}{2r}k_1x^2\bigg)^{-r}e^{\frac{2(2-r)}{3}k_2 t},&\;\;
r\neq0,2\\\\
\frac{2}{3}k_2t+\log(4k_2x+3k_1x^2+12\dot{x})-\frac{4(k_2x+3\dot{x})}
{(4k_2x+3k_1x^2+12\dot{x})},&\;\;r=2.
\end{array}\right.\label{the200}
\end{eqnarray}
Rewriting Eq.~(\ref{the200}) in the form (\ref{the02}), we get
\begin{eqnarray}
I=\left\{
\begin{array}{ll}
\bigg(\frac{(r-1)k_1^2}{4r^2}(xe^{\frac{k_2}{3}t})^4
+\frac {d}{dt}(xe^{\frac{k_2}{3}t})\bigg(\frac {d}{dt}(xe^{\frac{k_2}{3}t})
e^{\frac{k_2}{3}t}+\frac{k_1}{2}(xe^{\frac{k_2}{3}t})^2\bigg)e^{\frac{k_2}{3}t}
\bigg)\\\\
\qquad \qquad \times
\bigg(\frac {d}{dt}(xe^{\frac{k_2}{3}t})e^{\frac{k_2}{3}t}
+\frac{k_1(r-1)}{2r}(xe^{\frac{k_2}{3}t})^2\bigg)^{-r},&\;\;
r\neq0,2\\\\
\frac{4\frac {d}{dt}(xe^{\frac{k_2}{3}t})
e^{\frac{k_2}{3}t}}{k_1(xe^{\frac{k_2}{3}t})^2
+4\frac {d}{dt}(xe^{\frac{k_2}{3}t})e^{\frac{k_2}{3}t}}
-\log\bigg(k_1(xe^{\frac{k_2}{3}t})^2+4\frac {d}{dt}(xe^{\frac{k_2}{3}t})
e^{\frac{k_2}{3}t}\bigg),&\;\;r=2.
\end{array}\right.\label{the20}
\end{eqnarray}
Identifying the dependent and independent variables from (\ref{the20}) 
and the relations (\ref{the03}), we obtain the transformation
\begin{eqnarray} 
w = xe^{\frac{k_2}{3}t}, \quad
z = -\frac {3}{k_2}e^{-\frac{k_2}{3}t}. 
\label{the21}
\end{eqnarray}
In terms of the new variables, (\ref{the21}), the first integral $I$ given above, 
(\ref{the20}), can be written as
\begin{eqnarray}
I=\left\{
\begin{array}{ll}
\bigg(w'+\frac{(r-1)}{2r}k_1w^2\bigg)^{-r} \bigg[\frac{(r-1)}{4r^2}k_1^2w^4
+w'(w'+\frac{k_1}{2}w^2)\bigg],& r\neq0,2\\\\
\frac{4w'}{k_1w^2+4w'}
-\log(k_1w^2+4w'),&r=2.
\end{array}\right.\label{the22}
\end{eqnarray}
On the other hand the transformation (\ref{the21}) modifies the
equation (\ref{case04}) to the form
\begin{eqnarray}
w''+k_1ww'+\frac{(r-1)k_1^2}{2r^2}w^3=0,\;\; r\neq0\;\;
\mbox{and}\; '=\frac{d}{dz}.
\label{the24}
\end{eqnarray}

Finally, for the case $r=0$, we have an equation of the form (vide case $(ivb)$ in Table
$I$), $\ddot{x}+k_2\dot{x}+k_3x^3+\frac{2}{9}k_2^2x=0$, which is nothing but the
force-free Duffing oscillator equation. Again using the transformation 
(\ref{the21}),
the associated time dependent integral given in Table I can be rewritten as
\begin{eqnarray}
  \quad\quad 
I=\frac{w'^2}{2}+\frac{k_3}{4}w^4,
 &\;\;\;\; r=0.\label{the22c} 
\end{eqnarray}

Though it is difficult to integrate the above time independent first integrals,
(\ref{the22}), as 
they are in complicated forms, one can easily check that Eq.~(\ref{the22c}) 
$(r=0)$ can be integrated in terms of Jacobian elliptic 
function$\footnotesize^{14}$ and the case $r=1$ is already discussed as case $(ii)$ in 
this section. For the other cases one can give a Hamiltonian formulation as in
Sec. \ref{sec314} and write the corresponding Hamiltonian as 
\begin{eqnarray}
H=\left\{
\begin{array}{ll}
\bigg[\frac{\bigg((r-1)p\bigg)^{\frac{r-2}{r-1}}}{(r-2)}-p(
\frac{(r-1)}{2r}k_1w^2)\bigg],& r\neq0,1,2,\\ \\
\frac{k_1}{4}w^2p+\log(\frac{4k_1}{p}),& r=2\\\\
e^{p}+\frac{k_1}{2}w^2,&r=1\\\\
\frac{p^2}{2}+\frac{k_3}{4}w^4, & r=0
\end{array}\right.\label{the23e}
\end{eqnarray}
where
\begin{eqnarray}
p=\left\{
\begin{array}{ll}
\frac{1}{r-1}\bigg(\frac{(r-1)}{2r}k_1w^2
+w'\bigg)^{1-r},& r\neq0,1\\ \\
\log(w'),&r=1\\\\
w', & r=0.
\end{array}\right.\label{the23h}
\end{eqnarray}
Thus one is ensured of Liouville integrability of system (\ref{the24}) 
and so (\ref{case04}) for all values of $r$. Further, following the analysis in
the above subsection \ref{sec315}, one can make a canonical transformation (vide
Eqs.~(\ref{com05})-(\ref{com06})) to standard nonlinear oscillator equations.
\subsection{Summary of results for the $q=1$ case:}
\label{sec35}
To summarize the results obtained in this section, we have identified six 
integrable cases in Eq.~(\ref{int07}) among which four of
them were already known in the literature and the remaining two are new. In the
following, we tabulate all of them for convenience.
\subsubsection{Integrable equations already known in the literature}
\label{sec351}
\begin{eqnarray}
 (1) \qquad\;&&\ddot{x}+(k_1x+k_2)\dot{x}+\frac{k_1^2}{9}x^3
+\frac{k_1k_2}{3}x^2+\lambda_1x=0,\qquad \qquad \qquad \qquad
\qquad \qquad \qquad\nonumber (\ref {case01})\\
 (2) \qquad\;&&\ddot{x}+(k_1x+k_2)\dot{x}+\frac{k_1k_2}{3}x^2
+\frac{2k_2^2}{9}x=0,\qquad \qquad \qquad \qquad
\qquad \qquad \qquad \;\;\qquad \nonumber (\ref {case02a})\\
 (3) \qquad\;&&\ddot{x}+k_2\dot{x}+k_4x^2+\frac{6k_2^2}{25}x=0,
\qquad \qquad \qquad \qquad \qquad \qquad \qquad \qquad \qquad\quad\qquad\;
\nonumber (\ref {case03})\\
 (4)\qquad\;&&\ddot{x}+k_3x^3+k_4x^2+\lambda_1x=0.
\qquad \qquad \qquad \qquad \qquad \qquad \qquad \qquad \qquad\quad\qquad\;
\nonumber (\ref {cha4})
\end{eqnarray}
We note that the dynamics and certain transformation properties of 
Eq.~(\ref{case01}) have been studied in detail by three of the present authors
in Refs. 9 and 11 recently. In particular, we have shown that this 
equation admits certain unusual nonlinear dynamics$\footnotesize^{19}$. 
The dynamics of Eqs. (\ref{case02a}),(\ref{case03}) and (\ref{cha4})
can be found in Ref. 12.
\subsubsection{New integrable equations}
\label{sec352}
\begin{eqnarray}
 (1) \qquad\;&&\ddot{x}+k_1x\dot{x}+k_3x^3
+\lambda_1x=0,\qquad \qquad \qquad\qquad \qquad \qquad \qquad 
\qquad \quad\quad\;\;\;~\qquad\nonumber 
(\ref {cha3}) \\          
(2) \qquad\;&&\ddot{x}+(k_1x+k_2)\dot{x}+k_3x^3
+\frac{k_1k_2}{3}x^2+\frac{2k_2^2}{9}x=0,
\qquad \qquad \qquad\qquad \qquad \quad \qquad\;\;\nonumber (\ref {case04})
\end{eqnarray}
where $r^2k_3=\frac{(r-1)k_1^2}{2}$ and $k_1,k_2,\lambda_1$ and $r$ are 
arbitrary parameters. 
We note that (\ref{cha3}) includes the first equation of MEE hierarchy
(\ref{int05}) as a sub-case. Importantly, we showed that (\ref{cha3}) is a 
Hamiltonian system (see Eq.~(\ref{mlin012})) and so it is Liouville integrable.
Equation (\ref{case04}) can be transformed to the integrable Eq.~(\ref{the24}). 
Explicit general solution of certain special cases, namely,
$r=3$ or $\frac{3}{2}$ and $r=-1$ or $\frac{1}{2}$ are reported
in Ref. 20.

\section{Generalized force free DVP form of equations}
\label{sec4}
Let us now consider the case $q=2$ in Eq.~(\ref{int08}) or equivalently 
(\ref{int07b}), that is, 
\begin{eqnarray}            
\qquad \qquad\ddot{x}=-((k_1x^2+k_2)\dot{x}+k_3x^5+k_4x^3+\lambda_1x)\equiv\phi(x,\dot{x}).
\qquad \qquad \qquad\qquad\qquad \quad \;\;
\nonumber 
(\ref {int07b})
\end{eqnarray}
Interestingly Eq.~(\ref{int07b}) includes another class of physically important
nonlinear oscillators. For example, choosing $k_3=0$ one can get force-free 
Duffing-van der Pol oscillator equation. With the choice 
$k_2,\;k_4,\;\lambda_1=0$, it coincides with 
the second equation in the MEE hierarchy equation. Equation~(\ref{int07b}) with
the restriction $k_3=\frac{k_1^2}{16},\;k_4=\frac{k_1k_2}{4}$ and
$\lambda_1=(\omega_0^2+\frac{k_2^2}{4})$, has been
investigated in a different perspective in Ref. 4. However, the general
equation of the form (\ref{int07b}) has never been considered for integrability
test and so we perform the same here. 

To identify integrals of motion and the general solution of 
Eq.~(\ref{int07b}) we again seek the PS procedure. As the calculations are
similar to the $q=1$ case of Eq.~(\ref{int08}) which was carried out in the 
previous section, in the following, we give only the important steps.

\subsection{The case $I_t=0$}
\label{sec41}
By considering the same arguments given in Sec. \ref{sec311}, the null form $S$ can be
fixed easily in the form
\begin{equation}
S=-\frac{((k_1x^2+k_2)\dot{x}+k_3x^5+k_4x^3+\lambda_1x)}{\dot{x}}. 
\label{slin06}
\end{equation}
The respective $R$ equation becomes
\begin{eqnarray}          
 &&R_t+\dot{x}R_x-((k_1x^2+k_2)\dot{x}+k_3x^5+k_4x^3+\lambda_1x)R_{\dot{x}}  
\nonumber\\&&\qquad\qquad
=((k_1x^2+k_2)+\frac{((k_1x^2+k_2)\dot{x}+k_3x^5+k_4x^3+\lambda_1x)}
{\dot{x}})R.\label {slin03a}
\end{eqnarray}
To seek
a particular form for $R$ one may seek a suitable ansatz. We assume 
$R$ to be of the form (\ref{mlin08}) and investigate the system (\ref{slin03a})
as before. Following a similar procedure we find that a nontrivial particular solution for
(\ref{slin03a}) exists in the form
\begin{equation}
R=\frac{\dot{x}}{(\frac{(r-1)}{r}(\frac{k_1}{3}x^3+k_2x)+\dot{x})^r},
\label{slin11}
\end{equation}
where $r,k_1$ and $k_2$ are arbitrary parameters and the remaining parameters, 
$k_3,k_4$ and $\lambda_1$, are fixed by the relations 
\begin{eqnarray}
k_3=\frac{(r-1)}{3r^2}k_1^2,\quad 
k_4=\frac{4(r-1)}{3r^2}k_1k_2,
\quad \lambda_1=\frac{(r-1)}{r^2}k_2^2. 
\label{slin10}
\end{eqnarray} 
Further, we confirmed the compatibility of the functions $S$ and $R$ with the extra
constraint (\ref{lin04}) also. We note that unlike the earlier case, $q=1$, we do
not get a nontrivial solution for the parametric restriction $k_2,\;k_4=0$.
The above restrictions fix the Eq.~(\ref{int07b}) to the following specific forms:
\begin{subequations}
\begin{eqnarray}
  (ia) \;&&\ddot{x}+(k_1x^2+k_2)\dot{x}+\frac{(r-1)}{3r^2}k_1^2x^5
+\frac{4(r-1)k_1k_2}{3r^2}x^3+\frac{(r-1)k_2^2}{r^2}x=0,\;r\neq0
\label {ham01}\\
  (ib) \;&& \ddot{x}+k_3x^5+k_4x^3+\lambda_1 x=0,\;r=0
\label {ham01a}
\end{eqnarray}
\label{ham03}
\end{subequations}
Now substituting (\ref{slin06}) and (\ref{slin11}) into (\ref{met13}) and 
evaluating the integrals we obtain the first integrals in the form
\begin{subequations}
\begin{eqnarray}
  (ia) \quad&& I_1=
\bigg(\dot{x}+\frac{(r-1)}{r}(\frac{k_1}{3}x^3+k_2x)\bigg)^{-r}\nonumber\\
&&\qquad\quad \times
\bigg[\dot{x}(\dot{x}+\frac{k_1}{3}x^3+k_2x)
+\frac{(r-1)}{r^2}(\frac{k_1}{3}x^3+k_2x)^2
\bigg],\;\; r\neq0,2, \label {sintm03}\\
  (ib) \quad 
&&I_1=\frac{6\dot{x}}{(6\dot{x}+3k_2x+k_1x^3)}
-\log(6\dot{x}+3k_2x+k_1x^3),\;\; r=2,\label {sintm04}\\ 
  (ii) \quad 
&&I_1=\frac{\dot{x}^2}{2}+\frac{k_3}{6}x^6+\frac{k_4}{4}x^4
+\frac{\lambda_1}{2}x^2, \;\; r=0.\label {sintm06}
\end{eqnarray}
\label{slin12}
\end{subequations}

Further, as in the $q=1$ case in Sec.~\ref{sec314}, the integrals (\ref{slin12}) 
can be recast into the Hamiltonian form
\begin{subequations}
\begin{eqnarray}
(ia) \quad  
&&H=\bigg[\frac{\bigg((r-1)p\bigg)^{\frac{r-2}{r-1}}}{(r-2)}-
\frac{(r-1)}{r}p(\frac{k_1}{3}x^3+k_2x)\bigg],
\;\; r\neq0,1,2,\label{slin12d}\\
(ib) \quad 
&&H=\frac{k_2}{2}xp+\frac{k_1}{6}x^3p+\log(\frac{6}{p}),
\;\; r=2,\label{slin12e}\\
 (ic) \quad  
&& H=e^p+\frac{k_1}{3}x^3+k_2x,
\;\;r=1,\label{slin12f}\\ 
(ii) \quad 
&&H=\frac{p^2}{2}+\frac{k_3}{6}x^6+\frac{k_4}{4}x^4+\frac{\lambda_1}{2}x^2,
 \;\; r=0.\label{slin12g}
\end{eqnarray}
\label{slin012}
\end{subequations}
where the corresponding canonical momenta respectively  are
\begin{subequations} 
\begin{eqnarray}
(ia,b) \quad &&p=\frac{1}{(r-1)}\bigg(\dot{x}+\frac{(r-1)}{r}(\frac{k_1}{3}x^3
+k_2x)\bigg)^{(1-r)}, \;\; r\neq0,1,\\
\label{slin12h}
(ic) \quad && 
p=\log\dot{x},\;\;r=1,\label{slin12j}\\ 
(ii) \quad &&
p=\dot{x}, \;\; r=0.\label{slin12k}
\end{eqnarray}
\end{subequations}
Note that in the above the parameters $r,\;k_1,\;k_2,\;k_3$ and $\lambda_1$ are
all arbitrary. We also note here that unlike the $q=1$ case discussed in
Sec.~\ref{sec3}, so far we have
been unable to find suitable canonical transformations for the above Hamiltonian
systems so that the standard 'potential' equation results. The problem is being
further investigated.

\subsection{The case $I_t\neq0$}
\label{sec42}
Now let us study the case $I_t\neq0$. In this case $S$
has to be determined from Eq.~(\ref{lin02}), that is,
\begin{eqnarray}
&&S_t+\dot{x}S_x-((k_1x^2+k_2)\dot{x}+k_3x^5+k_4x^3+\lambda_1x) S_{\dot{x}} 
\nonumber\\
 && \qquad\qquad \qquad=(2k_1x\dot{x}+5k_3x^4+3k_4x^2+\lambda_1)-(k_1x^2+k_2)S+S^2.
   \label{slin05a}
\end{eqnarray}
As we did in the $q=1$ case of Eq.~(\ref{int08}) we proceed to solve 
Eq.~(\ref{slin05a}) with
the same form of ansatz (\ref{mlin05}). Doing so we find that Eq.~(\ref{slin05a}) 
admits
non-trivial forms of solutions for certain specific parametric restrictions. We
report both the parametric values and their respective forms of $S$ in Table II.

Now substituting the forms of $S$ into Eq.~(\ref{lin03}) and solving the
resultant equation we obtain the corresponding forms of $R$. Once $S$ and $R$
are determined then one has to verify the compatibility of this solution with
the extra constraint (\ref{lin04}). Then one
can substitute the null forms and integrating factors into (\ref{met13}) and 
construct the associated
integrals of motion. We report the integrating factors $(R)$ 
and time-dependent integrals of motion $(I)$ in Table II.

The remaining task is to derive the general solution and establish the complete
integrability of Eq.~(\ref{int07b}) for each parametric restriction. We again 
adopt the
procedure given in Sec. \ref{sec33} and transform the time dependent integrals into time
independent ones and integrate the latter and deduce the general solution. As
the procedure is exactly the same we provide only the results in the following.
\begin{sidewaystable}
\caption{Parametric restrictions, null forms $(S)$, integrating factors $(R)$ 
and time dependent integrals of motion $(I)$ of \\
\centerline{$\ddot{x}+(k_1x^2+k_2)\dot{x}+k_3x^5+k_4x^3+\lambda_1x=0$ (identified
with the assumed ansatz form of $S$ and $R$)}}
\small
\begin{tabular}{|p{.9cm}|p{4.2cm}|p{4cm}|p{4cm}|p{7cm}|}
\hline
Cases & Parametric restrictions & Null form $(S)$ & Integrating factor $(R)$ & 
Integrals of motion $(I)$\\
\hline
(i) & $k_3=\frac{k_1^2}{16},\;k_4=\frac{k_1k_2}{4}$ &&&
$(a)\;I=e^{\mp\omega t}
\left(\frac{4\dot{x}+2(k_2\pm\omega)x+k_1x^3}
{4\dot{x}+2(k_2\mp\omega)x+k_1x^3}\right),$ \\
&&$ \displaystyle{\frac{\frac{k_1}{2}x^3-\dot{x}}{x}}$ &
$\displaystyle{\frac{xe^{\mp\omega t}}
{(\dot{x}-\frac{(k_2\pm\omega)}{2}x+\frac{k_1}{4}x^3)^2}}$&
$\;\;\;\;\;\;\;\;\;\;\;\;\;k_2,\lambda_1\neq0,\;\omega=(k_2^2-4\lambda_1)^{\frac{1}{2}}$
\\
&($k_1,\;k_2,\;\lambda_1:\;$arbitrary)& &&
$(b)\;I=-t+\frac{x}{(\frac{k_2}{2}x+\frac{k_1x^3}{4}+\dot{x})},
\;\;\;\;\;\;\; k_2^2=4\lambda_1$\\
\hline
(ii) & $k_3=0,k_4=\frac{k_1}{6}(k_2\pm\omega),$&
$\displaystyle{\frac{1}{2}(k_2\mp\omega)+k_1x^2}$  &
$\displaystyle{e^{\frac{(k_2\pm\omega)}{2}t}}$  &
$I=\bigg(\dot{x}+\frac{k_2\mp\omega}{2}x
+\frac{k_1}{3}x^3\bigg)
e^{(\frac{k_2\pm\omega}{2})t},$  \\
&($k_1,\;k_2,\;\lambda_1:\;$arbitrary)
& &&$\;\;\;\;\;\;\;\;\;\omega=(k_2^2-4\lambda_1)^{\frac{1}{2}}$\\
\hline
(iii) & $k_1,\;k_3=0,\;\lambda_1=\frac{2k_2^2}{9}$ & 
$\displaystyle{\bigg(\frac{\frac{k_2 }{3}\dot{x}+\frac{k_2^2}{9} x+k_4x^3}
{\dot{x}+\frac{k_2}{3}x}\bigg)}$ &
$\displaystyle{{(\dot{x}+\frac{k_2}{3}x)}e^{\frac{4}{3}k_2 t}}$ &
$I=e^{\frac{4}{3}k_2 t} \bigg[\frac{\dot{x}^2}{2}
+\frac{k_2}{3} x\dot{x}+\frac{k_2^2 }{18}x^2+\frac{k_4}{4}x^4\bigg]$ \\
&($k_2,\;k_4:\;$arbitrary)&&&\\
\hline
(iva) & $k_3=\frac{(r-1)k_1^2}{3r^2},\;k_4=\frac{k_1k_2}{4},$ &
& &
$I=\bigg(\frac{k_3}{3}x^6
+(\dot{x}+\frac{k_2}{4}x)(\dot{x}+\frac{k_2}{4}x+\frac{k_1}{3}x^3)\bigg)$
\\
&$\lambda_1=\frac{3k_2^2}{16},\;r\neq0$&
$\displaystyle{\frac{k_2}{4}+k_1x^2+\frac {4k_3x^5}{(4\dot{x}+k_2x)}}$ &
$\displaystyle{\frac{(k_2x+4\dot{x})e^{\frac{3(2-r)}{4}k_2t}}
{(\frac{k_2}{4}x+rk_3x^3+\dot{x})^{r}}}$&
$\;\;\;\times \bigg(\dot{x}+\frac{k_2}{4}x
+rk_3x^3\bigg)^{-r}e^{\frac{3(2-r)}{4}k_2 t}, \;\;r\neq2$ 
\\
&($k_1,\;k_2,\;r:\;$arbitrary)
&&&$I=\frac{3}{4}k_2t+\log(6k_2x+4k_1x^3+24\dot{x})$
\\
&&&&$\;\;\;\;- \displaystyle{\frac{6(k_2x+4\dot{x})}
{(6k_2x+4k_1x^3+24\dot{x})}},\;\;\;\;\;\;\;\;\;\;\;\;\;\;r=2$\\
\hline
(ivb) & $k_1=0,\;k_4=0,$ &
$\displaystyle{\bigg(\frac{\frac{k_2}{4}\dot{x}+\frac{k_2^2}{16}x+k_3x^5}
{\dot{x}+\frac{k_2}{4}x}\bigg)}$ &
$\displaystyle{e^{\frac{3k_2}{2}t}(\dot{x}+\frac{k_2}{4}x)}$ &
$I=e^{\frac{3k_2}{2}t}\bigg(\frac{\dot{x}^2}{2}
+\frac{k_2}{4}x\dot{x}+\frac{k_2^2}{32}x^2+\frac{k_3}{6}x^6\bigg)$ \\
&$\lambda_1=\frac{3k_2^2}{16},r=0$&&&\\
&($k_2,\;k_3:\;$arbitrary)
&&&
\\
\hline
\end{tabular}
\end{sidewaystable}

$\bf{Case\;(ia)}$:$\;\underline{k_3=\frac{k_1^2}{16},\;k_4=\frac{k_1k_2}{4},
\;k_1,\;k_2\;\mbox{and}\;\lambda_1:\; \mbox{arbitrary}}$:\\

Substituting the parametric restrictions given above in Eq.~(\ref{int07b}),
we get 
\begin{eqnarray}
\ddot{x}+(k_1x^2+k_2)\dot{x}+\frac{k_1^2}{16}x^5
+\frac{k_1k_2}{4}x^3+\lambda_1x=0.
\label {tcase01}
\end{eqnarray}
We observed that the first integral of this case $(i)$ (see Table II), when
rewritten, is nothing but the Bernoulli equation which can be integrated
strightforwardly$\footnotesize^{21}$ and it leads to 
the general solution of the form
\begin{eqnarray} 
 \quad x(t)=\bigg(\frac{8k_2\lambda_1(e^{\omega t}-I_1)^2}
{I_1^2k_1k_2(-k_2+\omega)
-e^{2\omega t}k_1k_2(k_2+\omega)
+8I_2k_2\lambda_1 e^{(k_2+\omega) t}
+8I_1k_1\lambda_1 e^{\omega t}}\bigg)^{\frac{1}{2}},
\label{sthe07a}
\end{eqnarray}
where $\omega=\sqrt{k_2^2-4\lambda_1}$. A sub-case of
the Eq.~(\ref{tcase01}), namely, $k_2^2<4\lambda_1$ has been
studied by Smith$\footnotesize^{4,22}$ who showed that the corresponding
equation of motion admits a damped oscillatory form of solution, namely,
\begin{eqnarray} 
 \quad x(t)=\frac{Acos(\omega_0t+\delta)}{\bigg(e^{k_2t}-\frac{k_1A}{4k_2}
+\frac{k_1A}{4(k_2^2+4\omega_0^2)}\bigg(2\omega_0sin2(\omega_0t+\delta)
-k_2cos2(\omega_0t+\delta)\bigg)\bigg)^{\frac{1}{2}}},
\label{sthe07c}
\end{eqnarray}
where $\omega_0=\frac{1}{2}\sqrt{4\lambda_1+k_2^2}$ and $\delta,\;A$ are arbitrary
constants. 

On the other hand for $k_2^2> 4\lambda_1$, the solution (\ref{sthe07a}) becomes 
dissipative type having a front-like structure. In particular, for 
$\lambda_1=0$ we get a solution of the form
\begin{eqnarray} 
 \qquad\qquad x(t)=\bigg(\frac{2\sqrt{k_2}(I_1e^{k_2t}-1)}
{(-k_1+2k_1I_1e^{k_2t}(2+k_2I_1te^{k_2t})+4k_2I_2e^{2k_2t})^{\frac{1}{2}}}\bigg).
\label{sthe07b}
\end{eqnarray}\\

$\bf{Case\;(ib)}$:$\;\underline{k_3=\frac{k_1^2}{16},\;k_4=\frac{k_1k_2}{4},
\;k_2^2=4\lambda_1,\; k_1\;\mbox{and}\;k_2:\; \mbox{arbitrary}}$:\\

In this case we get the general solution of the form from (\ref{sthe07b}) as
\begin{eqnarray}
 \qquad\qquad x(t)=\bigg(\frac{2(I_1+t)^2}{2e^{k_2t}I_2-\frac{k_1}{k_2^3}
 (2+I_1^2k_2^2+2k_2t+k_2^2t^2+2I_1k_2(1+k_2t))}\bigg)^{\frac{1}{2}}.
\label{sthe07f}
\end{eqnarray}
One may note that a sub-case of this equation, namely, $k_2=\lambda_1=0$ leads
us to the second equation in the MEE hierarchy (\ref{int05}) and the
corresponding solution follows from Eq.~(\ref{sthe07f}) as
\begin{eqnarray}
x(t)=\sqrt{6}\bigg(\frac{(I_1+t)^2}{6I_2+k_1t(3I_1^2+3I_1t+t^2)}\bigg)^
{\frac{1}{2}}.
\label{sthe07e}
\end{eqnarray}
This form exactly coincides with the solution (\ref{hie05}) for $l=2$.\\

$\bf{Case\;(ii)}$: $\;\underline{k_3=0,\;k_4=\frac{k_1}{6}(k_2\pm\sqrt{k_2^2
-4\lambda_1}),\;k_1,\;k_2\;\mbox{and}\;\lambda_1:\; \mbox{arbitrary}}$:\\ 

The repective equation of motion and the first integral are (see Table II)
\begin{eqnarray}
\ddot{x}+(k_1x^2+k_2)\dot{x}+\frac{k_1}{6}(k_2\pm\sqrt{k_2^2
-4\lambda_1})x^3+\lambda_1x=0,
\label {tcase02}
\end{eqnarray}
and 
\begin{eqnarray} 
I=\bigg(\dot{x}+\frac{k_2\mp\sqrt{k_2^2-4\lambda_1}}{2}x
+\frac{k_1}{3}x^3\bigg)
e^{\frac{k_2\pm\sqrt{k_2^2-4\lambda_1}}{2}t}.
\label {tthe08}
\end{eqnarray}
Eq.~(\ref{tthe08}) can be rewritten as an Abel's equation in the form
\begin{eqnarray} 
\dot{x}=Ie^{(\frac{-k_2\mp\sqrt{k_2^2-4\lambda_1}}{2})t}
-\bigg(\frac{k_2\mp\sqrt{k_2^2-4\lambda_1}}{2}\bigg)x-\frac{k_1}{3}x^3.
\label {trthe08}
\end{eqnarray}
It is not clear whether Eq.~(\ref{trthe08}) can be explicitly integrated in 
general. However, for the special case $\lambda_1=\frac{3}{16}k_2^2$ it can be
integrated as follows.

The restriction $\lambda_1=\frac{2k_2^2}{9}$ fixes the equation of motion
(\ref{tcase02}) and the first integral (\ref{tthe08}) in the forms
\begin{eqnarray}
\ddot{x}+(k_1x^2+k_2)\dot{x}+\frac{k_1k_2}{4}x^3+\frac{3k_2^2}{16}x=0,
\label {tcase02a}
\end{eqnarray}
and
\begin{eqnarray} 
I=\bigg(\dot{x}+\frac{k_2}{4}x+\frac{k_1}{3}x^3\bigg)
e^{\frac{3k_2}{4}t},
\label {ttthe08}
\end{eqnarray}
respectively.

Now following our procedure given in Sec.~3.3 one arrives at the general 
solution$\footnotesize^1$ as  
\begin{eqnarray} 
z+z_0=-\frac {a}{3I}\left[\frac{1}{2}
\log\left(\frac{(w-a)^2}{w^2+aw+a^2}\right)
+\sqrt{3}\,\mbox{arctan}\left(\frac{-w\sqrt{3}}{2a+w}\right)\right], 
\label{sthe11a}
\end{eqnarray}
with $w = xe^{\frac{k_2}{4}t},\;z = -\frac {2}{k_2}e^{-\frac{k_2}{2}t}$ and 
$a=\sqrt[3]{\frac {3I}{k_1}}$ and $z_0$ is the second integration constant.  
Rewriting $w$ and $z$ in terms of old variables one can get the explicit 
solution.\\
 
$\bf{Case\;(iii)}$: $\;\underline{k_1,\;k_3=0,\;\lambda_1=\frac{2k_2^2}{9},
\;k_2\;\mbox{and}\;k_4:\; \mbox{arbitrary}}$:\\
 
The parametric restrictions given above fix the equation of motion
(\ref{int07b}) to the force-free Duffing oscillator, namely, 
$\ddot{x}+k_2\dot{x}+k_4x^3+\frac{2k_2^2}{9}x=0$. We 
have already discussed the general solution of this equation in Sec.~\ref{sec3} (vide
case $(iv)$).\\

$\bf{Case\;(iv)}$:$\;\underline{k_3=\frac{(r-1)k_1^2}{3r^3},
\;k_4=\frac{k_1k_2}{4},\;\lambda_1=\frac{3k_2^2}{16},
\;k_1,\;k_2\;\mbox{and}\;r:\; \mbox{arbitrary}}$:\\

The equation of motion turns out to be
\begin{eqnarray}
\ddot{x}+(k_1x^2+k_2)\dot{x}+\frac{(r-1)k_1^2}{3r^2}x^5+\frac{k_1k_2}{4}x^3
+\frac{3k_2^2}{16}x=0,\;\;r\neq0.
\label {tcase04}
\end{eqnarray}
Rewriting the associated first integral $I$, given in Case $(iv)$ in Table II,
in the form (\ref{the02}), we get
\begin{eqnarray}
I=\left\{
\begin{array}{ll}
\bigg(\frac{(r-1)k_1^2}{9r^2}(xe^{\frac{k_2}{4}t})^6
+\frac {d}{dt}(xe^{\frac{k_2}{4}t})\bigg(\frac {d}{dt}(xe^{\frac{k_2}{4}t})
e^{\frac{k_2}{2}t}+\frac{k_1}{3}(xe^{\frac{k_2}{4}t})^3\bigg)e^{\frac{k_2}{2}t}
\bigg) \\\\
\qquad\qquad \times 
\bigg(\frac {d}{dt}(xe^{\frac{k_2}{4}t})e^{\frac{k_2}{2}t}
+\frac{k_1(r-1)}{3r}(xe^{\frac{k_2}{4}t})^3\bigg)^{-r},&
 r\neq0,2,\\\\
\frac{6\frac {d}{dt}(xe^{\frac{k_2}{4}t})
e^{\frac{k_2}{2}t}}{k_1(xe^{\frac{k_2}{4}t})^3
+6\frac {d}{dt}(xe^{\frac{k_2}{4}t})e^{\frac{k_2}{2}t}}
-\log(k_1(xe^{\frac{k_2}{4}t})^3+6\frac {d}{dt}(xe^{\frac{k_2}{4}t})
e^{\frac{k_2}{2}t}),&r=2\\\\
\frac{1}{2}\bigg(\frac {d}{dt}(xe^{\frac{k_2}{4}t})\bigg)^2e^{k_2t}
+\frac{k_3}{6}(xe^{\frac{k_2}{4}t})^6,& r=0
\end{array}\right. \label{tthe20}
\end{eqnarray}
and identifying the dependent and independent variables from (\ref{tthe20}) 
and the relations (\ref{the03}), we obtain the transformation
\begin{eqnarray} 
w = xe^{\frac{k_2}{4}t}, \quad
z = -\frac {2}{k_2}e^{-\frac{k_2}{2}t}. 
\label{sthe21}
\end{eqnarray}
In terms of the new variables (\ref{sthe21}) the first integral $I$ given above,
(\ref{tthe20}) can be written as
\begin{eqnarray}
I=\left\{
\begin{array}{ll}
\bigg(w'+\frac{(r-1)}{3r}k_1w^3\bigg)^{-r} \bigg[w'(w'+\frac{k_1}{3}w^3)
+\frac{(r-1)}{9r^2}k_1^2w^6
\bigg],&r\neq0,2\\\\
\frac{6w'}{k_1w^3+6w'}-\log(k_1w^3+6w'),&r=2,\\\\
\frac{w'^2}{2}+\frac{k_3}{6}w^6,&  r=0. 
\end{array}\right. \label{sthe22}
\end{eqnarray}

On the other hand substituting the transformation (\ref{sthe21}) into the 
equation of motion (\ref{tcase04}) we get 
\begin{eqnarray}
w''+k_1w^2w'+\frac{(r-1)k_1^2}{3r^2}w^5=0,\;\;r\neq0, \qquad '=\frac{d}{dz}.
\label{sthe24}
\end{eqnarray}
In the case $r=0$, we have an equation of the form (vide case $(ivb)$ in 
Table II)
\begin{eqnarray}
\ddot{x}+k_2\dot{x}+k_3x^5+\frac{3k_2^2}{16}x=0.
\label {ttcase04}
\end{eqnarray}
We note that the Eq.~(\ref{sthe24}) is the $l=2$ case of Eq.~(\ref{int05}). As
we mentioned in the introduction the general solution of this equation can be
obtained only for certain specific choices, namely,
$\frac{(r-1)k_1^2}{3r^2}=\frac{1}{16}$. This in turn gives $r=4k_1$ or
$\frac{4}{3}k_1$. The respective solutions for these values of $r$ of 
Eq.~(\ref{sthe24}) can be fixed from Eq.~(\ref{hie05}) 
with $l=2$. The other cases do not seem to be 
amenable to explicit integration. However, all of them can be recast in the 
Hamiltonian form as we see below.

As the first integrals (\ref{sthe22}) are now `time' independent ones, one can 
give a Hamiltonian
formalism for all the integrals (\ref{sthe22}) by following the ideas given in
Sec.~\ref{sec314}. Doing so we obtain
\begin{eqnarray}
H=\left\{
\begin{array}{ll}
\bigg[\frac{\bigg((r-1)p\bigg)^{\frac{r-2}{r-1}}}{(r-2)}-
\frac{(r-1)}{3r}k_1w^3p\bigg],& r\neq0,1,2,\\ \\
\frac{k_1}{6}w^3p+\log(\frac{6}{p}),& r=2\\\\
e^p+\frac{k_1}{3}w^3,&r=1\\\\
\frac{p^2}{2}+\frac{k_3}{6}w^6, & r=0
\end{array}\right.\label{sthe23f}
\end{eqnarray}
where 
\begin{eqnarray}
p=\left\{
\begin{array}{ll}
\frac{1}{(r-1)}\bigg(w'+\frac{(r-1)}{3r}k_1w^3
\bigg)^{(1-r)},& r\neq0,1\\ \\
\log{w'},&r=1\\\\
\frac{p^2}{2}+\frac{k_3}{6}w^6, & r=0.
\end{array}\right.\label{sthe23e}
\end{eqnarray}

In this sense these cases may be considered as Liouville integrable systems. 
Finally, for $r=0$ case in Eq.~(\ref{sthe22}) can be integrated in terms of 
Jacobian elliptic function (see for example in Ref. 23). Again, here, we have not
been able to identify canonical transformations which can lead to the
identification of suitable 'potential' equations. 
\subsection{Summary of results in $q=2$ case:}
\label{sec43}
To summarize the results obtained for the $q=2$ case, we have identified six 
integrable cases in Eq.~(\ref{int07b}) among which three of
them were already known in the literature and the remaining three are new. In the
following, we tabulate both of them.
\subsubsection{Integrable equations already known in the literature}
\label{sec431}
\begin{eqnarray}
  (1)\quad&&\ddot{x}+(k_1x^2+k_2)\dot{x}+\frac{k_1k_2}{4}x^3
+\frac{3k_2^2}{16}x=0,\qquad \qquad \qquad \qquad \qquad \qquad 
\qquad\qquad \quad\nonumber (\ref {tcase02a})\\
  (2)\quad&&\ddot{x}+k_2\dot{x}+k_3x^3+\frac{2k_2^2}{9}x=0,
\label {lcom6b}\\
  (3) \quad&& \ddot{x}+k_3x^5+k_4x^3+\lambda_1 x=0.
\qquad \qquad \qquad \qquad \qquad \qquad 
\qquad\qquad\qquad\qquad\qquad \;\;\nonumber (\ref {ham01a})
\end{eqnarray}
We mention that Eq.~(\ref{tcase02a}) is nothing but the force-free DVP whose 
integrability is established in Ref. 1 and Eq.~(\ref{lcom6b}) is 
nothing but the force-free Duffing oscillator$\footnotesize^{12,14}$.
\subsubsection{New integrable equations}
\label{sec432}
\begin{eqnarray}
  (1)\quad &&\ddot{x}+(k_1x^2+k_2)\dot{x}+k_3x^5
+\frac{4(r-1)k_1k_2}{3r^2}x^3+\frac{(r-1)k_2^2}{r^2}x=0,\;r\neq0
\qquad \qquad \quad \; \nonumber (\ref {ham01})\\
  (2)\quad &&\ddot{x}+(k_1x^2+k_2)\dot{x}+\frac{k_1^2}{16}x^5
+\frac{k_1k_2}{4}x^3+\lambda_1x=0,
\qquad \qquad \qquad \qquad \qquad \qquad\quad\;\;\;\;\;\;\nonumber (\ref {tcase01})\\
  (3)\quad&&\ddot{x}+(k_1x^2+k_2)\dot{x}+k_3x^5
+\frac{k_1k_2}{4}x^3+\frac{3k_2^2}{16}x=0,
\qquad \qquad \qquad \qquad \qquad \qquad\quad\;\;\; \nonumber (\ref {tcase04})
\end{eqnarray}
where $r^2k_3=\frac{(r-1)k_1^2}{3}$ and $k_1,k_2,\lambda_1$ and $r$ are 
arbitrary parameters.
We proved that Eq.~(\ref{ham01}) is Liouville integrable one. As far
as equation (\ref{tcase01}) is concerned we derived the general solution for 
arbitrary values of $k_1,\;k_2$ and $\lambda_1$. 
Finally, for Eq.~(\ref{tcase04}) though we identified only one time 
dependent integral, we have demonstrated that it can be transformed into time 
independent Hamiltonian and thereby ensuring its Liouville integrability.

\section{Extended Prelle-Singer method to generalized Eq.~(\ref{int08})}
\label{sec5}
One can investigate the integrability properties of Eq.~(\ref{int08}) by 
considering the cases $q=3,4,5,\ldots,$ one by one and classify the integrable
equations. Since the 
procedure and the mathematical techniques in exploring the integrating factors
$(R)$, null forms $(S)$, first integrals $(I)$ and general solution are the same in
each case we do not consider each case in detail. We straightaway move to
the case $q=arbitrary$, that is, $q\in R$ and not necessarily an integer, and 
present the outcome of our investigations.

As we did earlier, we consider the cases $I_t=0$ and $I_t\neq0$ separately for
the choice $q=arbitrary$ also.
First let us consider the case $I_t=0$. 
\subsection{The case $I_t=0$}
\label{sec51}
By considering the same arguments given in Sec. 3.1.1 the null form $S$ and the
integrating factor $R$ can be fixed easily in the form
\begin{eqnarray}
&&S=-\frac{((k_1x^q+k_2)\dot{x}+k_3x^{2q+1}+k_4x^{1+q}+\lambda_1x)}{\dot{x}},
\nonumber\\
&&R=\frac{\dot{x}}{(\frac{(r-1)}{r}(\frac{k_1}{(q+1)}x^{q+1}+k_2x)+\dot{x})^r},
\label{qlin11}
\end{eqnarray}
where $k_1$ and $k_2$ are arbitrary and the remaining parameters, $k_3,k_4$ and 
$\lambda_1$, are related to the parameters $k_1$ and $k_2$ through the 
relations 
\begin{eqnarray}
 k_3=\frac{(r-1)}{r^2}(q+1)\hat{k}_1^2,\quad
k_4=\frac{(r-1)}{r^2}(q+2)\hat{k}_1k_2,\quad
\lambda_1=\frac{(r-1)}{r^2}k_2^2, 
\label{qlin10}
\end{eqnarray}
where $\hat{k}_1=\frac{k_1}{(q+1)}$. The above restrictions fix Eq.~(\ref{int08}) 
to the following specific forms:
\begin{eqnarray}
\qquad\qquad \qquad (ia) \quad&& \ddot{x}+((q+1)\hat{k}_1x^q+k_2)\dot{x}
+\frac{(r-1)}{r^2}[(q+1)\hat{k}_1^2x^{2q+1}
\nonumber\\&&\qquad \qquad\qquad\qquad\qquad
+(q+2)\hat{k}_1k_2x^{q+1}+k_2^2x]=0,\;r\neq0
\qquad \quad\nonumber (\ref {ncom9})\\ 
\qquad\qquad  (ib) \quad && \ddot{x}+k_3x^{2q+1}+k_4x^{q+1}+\lambda_1 x=0,\;r=0.
\label {qham01a}
\end{eqnarray}
 
Now substituting (\ref{qlin11}) into (\ref{met13}) and 
evaluating the integrals we obtain the first integrals of the form
\begin{subequations}
\begin{eqnarray}
  (ia) \;&& I_1=
\bigg(\dot{x}+\frac{(r-1)}{r}(\hat{k}_1x^{q+1}+k_2x)\bigg)^{-r}\nonumber\\
&&\qquad\quad\times
\bigg[\dot{x}(\dot{x}+\hat{k}_1x^{q+1}+k_2x)
+\frac{(r-1)}{r^2}(\hat{k}_1x^{q+1}+k_2x)^2
\bigg],\;\; r\neq0,2, 
\label {qintm03}\\
  (ib) \; &&
I_1=\frac{\dot{x}}{(\dot{x}+\frac{k_2}{2}x+\frac{\hat{k}_1}{2}x^{q+1})}
-\log(\dot{x}+\frac{k_2}{2}x+\frac{\hat{k}_1}{2}x^{q+1}),\;\; r=2,\label {qintm04}\\ 
  (ii) \; &&
I_1=\frac{\dot{x}^2}{2}+\frac{k_3}{2(q+1)}x^{2(q+1)}+\frac{k_4}{(q+2)}x^{q+2}
+\frac{\lambda_1}{2}x^2, \;\; r=0.\label {qintm06}
\end{eqnarray}
\label{qlin12}
\end{subequations}

Further, using the above forms of the first integrals, one can show that the 
equation of motion (\ref{int08}), with the parametric restrictions 
(\ref{qlin10}), can also be derived from the Hamiltonians
\begin{subequations}
\begin{eqnarray}
 (ia) \quad && 
H=\bigg[\frac{\bigg((r-1)p\bigg)^{\frac{r-2}{r-1}}}{(r-2)}-
\frac{(r-1)}{r}p(\hat{k}_1x^{q+1}+k_2x)\bigg],
\;\; r\neq0,1,2,\label{qlin12d}\\
(ib) \quad &&
H=\frac{k_2}{2}xp+\frac{\hat{k}_1}{2}x^{q+1}p+\log(\frac{2(q+1)}{p}),
\;\; r=2\label{qlin12e}\\
(ic) \quad  &&
H=e^p+\hat{k}_1x^{q+1}+k_2x,
\;\;r=1,\label{qlin12f}\\ 
(ii) \quad &&
H=\frac{p^2}{2}+\frac{k_3}{2(q+1)}x^{2(q+1)}+\frac{k_4}{(q+1)}x^{q+1}
+\frac{\lambda_1}{2}x^2,
 \;\; r=0,\label{qlin12g}
\end{eqnarray}
\label{qlin012}
\end{subequations}
where the corresponding canonical momenta respectively  are
\begin{subequations} 
\begin{eqnarray}
(ia,b) \quad &&p=\frac{1}{(r-1)}
\bigg(\dot{x}+\frac{(r-1)}{r}(\hat{k}_1x^{q+1}+k_2x)\bigg)^{(1-r)},
 \;\; r\neq0,1,\\
\label{qlin12h}
 (ic) \quad && 
p=\log\dot{x},\;\;r=1,\label{qlin12j}\\ 
 (ii) \quad &&
p=\dot{x}, \;\; r=0.\label{qlin12k}
\end{eqnarray}
\end{subequations}

With the above Hamiltonian formulation, for the parametric set (\ref{qlin10}), the
integrability of the associated equation of motion is assured for these parametric 
cases through Liouville theorem.

\subsection{The case $I_t\neq0$}
\label{sec52}
We use the same ansatz and ideas which we followed for the $q=1$ and
$q=2$ cases to determine the forms of $S$ and $R$. As the procedure is 
exactly the same as in the earlier cases we present the parametric 
restrictions and the respective form of expressions of the integrating 
factors, null forms and integrals of motions in Table III without further
discussion.
\begin{sidewaystable}
\caption{Parametric restrictions, null forms $(S)$, integrating factors $(R)$ 
and time dependent integrals of motion $(I)$ of\\
\centerline{$\ddot{x}+(k_1x^q+k_2)\dot{x}+k_3x^{2q+1}+k_4x^{q+1}+\lambda_1x=0$
(identified
with the assumed ansatz form of $S$ and $R$)}}
\small
\begin{tabular}{|p{.9cm}|p{4.2cm}|p{4cm}|p{4.6cm}|p{8cm}|}
\hline
Cases & Parametric restrictions & Null form $(S)$ & Integrating factor $(R)$ & 
Integrals of motion $(I)$  \\
\hline
(i) & $k_3=\frac{k_1^2}{(q+2)^2},\;k_4=\frac{k_1k_2}{(q+2)}$ &
 & 
 &
$(a)\;I=e^{\mp\omega t}
\left(\frac{\dot{x}-\frac{(-k_2\mp\omega)}{2}x+\frac{k_1}{q+2}x^{q+1}}
{\dot{x}-\frac{(-k_2\pm\omega)}{2}x+\frac{k_1}{q+2}x^{q+1}}\right),
 $ \\
&&
$ \displaystyle{\frac{(\frac{qk_1}{(q+2)}x^{q+1}-\dot{x})}{x}}$ &
$\displaystyle{\frac{xe^{\mp\omega t}}
{(\dot{x}-\frac{(k_2\pm\omega)}{2}x+\frac{k_1}{(q+2)}x^{q+1})^2}}$&
$\;\;\;\;\;\;\;\;\;\;k_2,\lambda_1\neq0,\;\omega=(k_2^2-4\lambda_1)^{\frac{1}{2}}$\\
&($k_1,\;k_2,\;\lambda_1:\;$arbitrary)& &&
$(b)\;I=-t+\frac{x}{(\frac{k_2}{2}x+\frac{k_1x^{q+1}}{q+2}+\dot{x})},
\;\;\;\;\;\;\;\; k_2^2=4\lambda_1$\\
&&&&\\
\hline
(ii) &$k_4=\frac{k_1(k_2\pm\omega)}{2(q+1)},\;k_3=0,$&
$\displaystyle{\frac{1}{2}(k_2\mp\omega)+k_1x^q},$  &
$\displaystyle{e^{\frac{(k_2\pm\omega)}{2}t}}$  &
$I=\bigg(\dot{x}+\frac{k_2\mp\omega}{2}x
+\frac{k_1}{(q+1)}x^{q+1}\bigg)
e^{(\frac{k_2\pm\omega}{2})t},$  \\
&$(k_1,\;k_2,\;\lambda_1:\;$arbitrary)& 
&&$\;\;\;\;\;\;\;\;\;\;\omega=(k_2^2-4\lambda_1)^{\frac{1}{2}}$\\
\hline
(iii) & $k_1,k_3=0,\lambda_1=\frac{2(q+2)k_2^2}{(q+4)^2}$ & 
$\displaystyle{\frac{\frac{2k_2 \dot{x}}{(q+4)}
+\frac{4k_2^2 x}{(q+4)^2}+k_4x^{q+1}}
{(\dot{x}+\frac{2k_2 x}{(q+4)})}}$ &
$\displaystyle{{(\dot{x}+\frac{2k_2 x}{(q+4)})}
e^{\frac{2(q+2)}{(q+4)}k_2 t}}$ &
$I=e^{\frac{2(q+2)}{(q+4)}k_2 t} \bigg[\frac{\dot{x}^2}{2}
+\frac{2k_2 x\dot{x}}{(q+4)}+\frac{2k_2^2x^2  }{(q+4)^2}
+\frac{k_4x^{q+2}}{(q+2)}\bigg]$ \\
&($k_2,\;k_4:\;$arbitrary)
&&&\\
\hline
(iv)a & $k_3=\frac{(r-1)k_1^2}{(q+1)r^2},$ &
& &
$I=\bigg(\frac{k_3x^{2(q+1)}}{(q+1)}+(\dot{x}+\frac{k_2x}{q+2})
(\dot{x}+\frac{k_2x}{q+2}+\frac{k_1x^{q+1}}{q+1})\bigg)$
\\
&$k_4=\frac{k_1k_2}{(q+2)},$
&$\frac{k_2}{(q+2)}+k_1x^{q}+\frac {k_3x^{2q+1}}
{(\dot{x}+\frac{k_2}{(q+2)}x)}$
&$\displaystyle{\frac{(k_2x+(q+2)\dot{x})e^{\frac{(q+1)(2-r)}{(q+2)}k_2t}}
{(\frac{k_2}{(q+2)}x+rk_3x^{q+1}+\dot{x})^{r}}}$
&
$\;\;\;\times \bigg(\frac{k_2}{(q+2)}x+rk_3x^{q+1}+\dot{x}\bigg)^{-r}
e^{\frac{(q+1)(2-r)}{(q+2)}k_2t}, r\neq2$ 
\\
&$\lambda_1=\frac{(q+1)k_2^2}{(q+2)^2},\;r\neq0$&&
&$I=\frac{q+1}{q+2}k_2t+\log(k_1x^{q+1}+2(q+1)(\dot{x}+\frac{k_2}{q+2}x))$
\\
&($k_1,\;k_2,\;r:\;$arbitrary)&&&$\;\;\;\;- \displaystyle{(\frac{2(q+1)
(\dot{x}+\frac{k_2}{q+2}x)}
{k_1x^{q+1}+2(q+1)(\dot{x}+\frac{k_2}{q+2}x)})},
\;\;\;\;\;\;\;\;r=2$\\
\hline
(iv)b & $k_1=0,\;k_4=0,$ &
$\displaystyle{\frac{k_2}{(q+2)}+\frac {k_3x^{2q+1}}
{(\dot{x}+\frac{k_2}{(q+2)}x)}}$ &
$\displaystyle{e^{\frac{(2q+2)k_2}{(q+2)}t}(\dot{x}+\frac{k_2}{(q+2)}x)}$ &
$I=\bigg(\frac{\dot{x}^2}{2}+\frac{k_2x\dot{x}}{(q+2)}
+\frac{k_2^2x^2}{2(q+2)^2}+\frac{k_3x^{2q+2}}{(2q+2)}\bigg)
e^{\frac{(2q+2)k_2}{(q+2)}t}$ \\
&$\lambda_1=\frac{(q+1)k_2^2}{(q+2)^2},r=0$&&&\\
&($k_2,\;k_3:\;$arbitrary)
&&&
\\
\hline
\end{tabular}
\end{sidewaystable}

Since we derived only one integral, which is also a time dependent one for each
parametric restriction, we need to integrate each one of them further and 
obtain the second 
integration constant in order to prove the complete integrability of each of
the 
cases reported in Table III. In the following we deduce the second integral 
and general solution by utilizing the proceduce given in Sec. \ref{sec33}.\\

$\bf{Case\;(ia)}$:$\;\underline{k_3=\frac{k_1^2}{(q+2)^2},
\;k_4=\frac{k_1k_2}{(q+2)},
\;k_1,\;k_2\;\mbox{and}\;\lambda_1:\; \mbox{arbitrary}}$:\\

We have an equation of the form
\begin{eqnarray} 
\qquad \qquad \ddot{x}+((q+2)\hat{k}_1x^q+k_2)\dot{x}+\hat{k}_1^2x^{2q+1}
+\hat{k}_1k_2x^{q+1}+\lambda_1x=0,
\qquad \qquad \qquad \qquad \quad\;     \nonumber (\ref {ncom7})
\end{eqnarray}
where $k_1=(q+2)\hat{k}_1$. The corresponding first integral given in Table 3 
is nothing but the Bernoulli equation 
which can be solved using the standard method$\footnotesize^{21}$. The general 
solution turns out to be
\begin{eqnarray} 
x(t)=\bigg(e^{\omega t}-I_1\bigg)
\bigg(e^{\frac{q}{2}(k_2+\omega)t}\bigg(I_2+\hat{k}_1q
\int{\bigg(\frac{e^{\omega t}-I_1}{e^{\frac{1}{2}(k_2+\omega)t}}\bigg)^q dt}\bigg)\bigg)
^{\frac{-1}{q}},
\label{qthe07a}
\end{eqnarray}
where $\omega=\sqrt{k_2^2-4\lambda_1}$. We note here that a sub-case
of the above, namely, $k_2^2< 4\lambda_1$, has been
studied by Smith$\footnotesize^4$ who had shown that the corresponding 
system admits the general solution of the form 
\begin{eqnarray} 
 x(t)=Acos(\omega_0t+\delta)e^{-\frac{k_2}{2}t}\bigg(1
+q\hat{k}_1A\int{e^{\frac{-qk_2}{2}t}
cos^q(\omega_0t+\delta) dt}\bigg)^{-\frac{1}{q}},
\label{qthe07c}
\end{eqnarray}
where $\omega_0=\frac{1}{2}\sqrt{4\lambda_1+k_2^2}$ and $\delta,\;A$ are 
arbitrary constants. For $k_2^2> 4\lambda_1$, the solution become a dissipative 
type/front-like structure. In particular, for $\lambda_1=0$ the general 
solution takes the form
\begin{eqnarray} 
 x(t)=\bigg(e^{k_2 t}I_1-1\bigg)
\bigg[e^{qk_2t}\bigg(I_2+\hat{k}_1q
\int{\bigg(I_1-e^{-k_2t}\bigg)^q dt}\bigg)\bigg]
^{-\frac{1}{q}}.
\label{qthe07b}
\end{eqnarray}\\

$\bf{Case\;(ib)}$:$\;\underline{k_3=\frac{k_1^2}{16},\;k_4=\frac{k_1k_2}{4},
\;k_2=4\lambda_1,\; k_1\;\mbox{and}\;k_2:\; \mbox{arbitrary}}$:\\
  
A general solution for this case can be fixed from (\ref{qthe07b}) as 
\begin{eqnarray}
 x(t)=(I_1+t)e^{-\frac{k_2}{2}t}\bigg(I_2+q\hat{k}_1
 \int{e^{-\frac{qk_2}{2}t}(I_1+t)^q dt}\bigg)^{-\frac{1}{q}}.
\label{qthe07f}
\end{eqnarray} 
On the other hand the general solution for 
the parametric choice $k_2,\;\lambda_1=0$ turns out to be
\begin{eqnarray}
x(t)=\bigg(\frac{(q+1)(I_1+t)^q}{\hat{k}_1q(I_1+t)^{q+1}
+(q+1)I_2}\bigg)^{\frac{1}{q}},
\label{qthe07e}
\end{eqnarray}
which exactly coincides with the result (\ref{hie05}) obtained by  
Feix et al.$\footnotesize^{3}$ for integer $q(=l)$ values.\\

$\bf{Case\;(ii)}$: $\;\underline{k_3=0,\;k_4=\frac{k_1}{2(q+1)}(k_2\pm\sqrt{k_2^2
-4\lambda_1}),\;k_1,\;k_2\;\mbox{and}\;\lambda_1:\; \mbox{arbitrary}}$:\\

The associated equation of motion and the first integral are (see Table III)
\begin{eqnarray}
\ddot{x}+((q+1)\hat{k}_1x^q+k_2)\dot{x}+\frac{\hat{k}_1}{2}(k_2\pm\sqrt{k_2^2
-4\lambda_1})x^{q+1}+\lambda_1x=0,
\label {qcase02}
\end{eqnarray}
and 
\begin{eqnarray} 
I=\bigg(\dot{x}+\frac{k_2\mp\sqrt{k_2^2-4\lambda_1}}{2}x
+\hat{k}_1x^{q+1}\bigg)
e^{(\frac{k_2\pm\sqrt{k_2^2-4\lambda_1}}{2})t},
\label {qqthe08}
\end{eqnarray}
where $k_1=(q+1)\hat{k}_1$. Like in the earlier cases, that is, $q=1$ and $q=2$,
we are able to integrate the first integral (\ref{qqthe08}) explicitly only for
a specific parametric restriction, namely, $\lambda_1=(q+1)\hat{k}_2^2$,
where $k_2=(q+2)\hat{k}_2$. In this case the equation of motion
(\ref{qcase02}) and the first integral, Eq.~(\ref{qqthe08}), can be recast in 
the form
\begin{eqnarray}
\qquad \qquad \ddot{x}+(k_1x^q+(q+2)\hat{k}_2)\dot{x}+k_1\hat{k}_2x^{q+1}+(q+1)\hat{k}_2^2 x=0,
\qquad \qquad \qquad\qquad \qquad\quad  \nonumber (\ref {ncom11a})
\end{eqnarray}
and   
\begin{eqnarray} 
I=\bigg(\dot{x}+\hat{k}_2 x+\hat{k}_1x^{q+1}\bigg)e^{(q+1)\hat{k}_2t},
\label {qqthe08a}
\end{eqnarray} 
respectively. Now comparing (\ref{qqthe08a}) with (\ref{the02}), we get
\begin{eqnarray} 
I=e^{q\hat{k}_2 t}\bigg(\frac {d}{dt}(xe^{\hat{k}_2 t})\bigg)
+\hat{k}_1(xe^{\hat{k}_2t})^{(q+1)}.
\label {qthe09}
\end{eqnarray}

Next identifying the dependent and independent variables from (\ref{qthe09}) 
using the relations (\ref{the03}), we obtain the transformation
\begin{eqnarray} 
w = xe^{\hat{k}_2t}, \quad
z = -\frac{1}{q\hat{k}_2}e^{-q\hat{k}_2 t}. 
\label{qthe10}
\end{eqnarray}
Using the transformation (\ref{qthe10}) the first integral (\ref{qthe09}) 
can be rewritten in the form
\begin{equation}
I = w'+\hat{k}_1w^{(q+1)}
\label{qthe11}
\end{equation}
which in turn leads to the solution by an integration, that is,
\begin{eqnarray} 
z-z_0 = \int\frac{dw}{I-\hat{k}_1w^{(q+1)}},
\label{qthe011a}
\end{eqnarray}
where $z_0$ is an arbitrary constant. Solving Eq.~(\ref{qthe011a}) we 
get$\footnotesize^{24}$
\begin{eqnarray}
  z-z_0=\frac{1}{Ig^{\frac{1}{(q+1)}}}\left\{
\begin{array}{ll}
\displaystyle{-\frac{2}{q+1}\sum_{i=0}^{\frac{q-1}{2}}P_i\cos\frac{2i}{q+1}\pi
+\frac{2}{q+1}\sum_{i=0}^{\frac{q-1}{2}}Q_i\sin\frac{2i}{q+1}\pi}\\
+\frac{1}{q+1}ln\frac{(1+w)}{(1-w)},\quad \mbox{q-a positive odd number},\\ 
\displaystyle{-\frac{2}{q+1}\sum_{i=0}^{\frac{q-2}{2}}R_i\cos\frac{2i+1}{q+1}\pi
+\frac{2}{q+1}\sum_{i=0}^{\frac{q-2}{2}}T_i\sin\frac{2i+1}{q+1}\pi}\\
+\frac{1}{q+1}ln(1+w),\quad \mbox{q-a positive even number},
\end{array}\right.  \label {rt04}
\end{eqnarray}
where $g=\frac{\hat{k_1}}{I}$ and 
\begin{eqnarray} 
P_i=\frac{1}{2}ln\bigg(w^2-2w\cos\frac{2i}{q+1}\pi+1\bigg),
\quad Q_i=arctan\bigg[\frac{w-\cos\frac{2i}{q+1}\pi}{\sin\frac{2i}{q+1}\pi}\bigg],
\nonumber\\
R_i=\frac{1}{2}ln\bigg(w^2+2w\cos\frac{2i+1}{q+1}\pi+1\bigg),
\quad T_i=arctan\bigg[\frac{w+\cos\frac{2i+1}{q+1}\pi}{\sin\frac{2i+1}{q+1}\pi}
\bigg].\nonumber
\end{eqnarray}
Rewriting $w$ and $z$ in terms of old variables one can get the explicit 
solution.\\

$\bf{Case\;(iii)}$: $\;\underline{k_1,k_3=0,\;
\lambda_1=\frac{2(q+2)k_2^2}{(q+4)^2},
\;k_2\;\mbox{and}\;k_4:\; \mbox{arbitrary}}$\\

The parametric choice given above fixes the equation of motion of the form
\begin{eqnarray}
\qquad \qquad\ddot{x}+(q+4)\hat{k}_2\dot{x}+k_4x^{(q+1)}+2(q+2)\hat{k}_2^2x=0,
\qquad \qquad \qquad\qquad\qquad \qquad\qquad\;\; \nonumber (\ref {ncom11})
\end{eqnarray}
where $k_2=(q+4)\hat{k}_2$. Rewriting the first integral $I$ given in 
Case $(iii)$ in Table III, in the form 
(\ref{the01}), we get 
\begin{eqnarray}
I=\frac{1}{2}\left(\dot{x}+2\hat{k}_2 x\right)^2 
e^{2(q+2)\hat{k}_2 t}
+\frac{k_4x^{(q+2)}}{(q+2)}e^{2(q+2)\hat{k}_2 t}. \label {qthe15}
\end{eqnarray}
Now splitting the first term in Eq.~(\ref{qthe15}) further in the form of 
(\ref{the02}),
\begin{eqnarray} 
I=\bigg[e^{q\hat{k}_2 t}\frac{d}{dt}\left(\frac{x}{\sqrt{2}}
e^{2\hat{k}_2 t}\right)\bigg]^2
+\frac{2^{(\frac{q+2}{2})}k_4}{(q+2)}\left(\frac{x}{\sqrt{2}} 
e^{2\hat{k}_2 t}\right)^{(q+2)} \label {qthe16}
\end{eqnarray}
and identifying the dependent and independent variables from (\ref{qthe16})
using the relations (\ref {the03}), we obtain the transformation
\begin{eqnarray} 
w=\frac{x}{\sqrt{2}}e^{2\hat{k}_2 t},\;\;\;\;
z=-\frac {1}{q\hat{k}_2}e^{-q\hat{k}_2 t}. \label {qthe17}
\end{eqnarray}
Using the transformation (\ref{qthe17}) the first integral (\ref{qthe15}) 
can be brought to the form
\begin{equation}
I = w'^2+\frac{2^{(\frac{q+2}{2})}k_4}{(q+2)}w^{(q+2)}.
\label{qthe17a}
\end{equation}
Separating the dependent and independent variables and integrating the resultant
equation we get
\begin{eqnarray} 
z-z_0 = \int\frac{dw}{\sqrt{I-\hat{k}_4w^{(q+2)}}},
\label{qthe17aa}
\end{eqnarray}
where $\hat{k}_4=\frac{2^{(\frac{q+2}{2})}}{(q+2)}k_4$ and $z_0$ is an arbitrary 
constant.\\

$\bf{Case\;(iv)}$:$\;\underline{k_3=\frac{(r-1)k_1^2}{(q+1)r^2},
\;k_4=\frac{k_1k_2}{(q+2)},\;\lambda_1=\frac{(q+1)k_2^2}{(q+2)^2},
\;k_1,\;k_2\;\mbox{and}\;r:\; \mbox{arbitrary}}$:\\

The equation of motion in this case turns out to be
\begin{eqnarray}
&& \qquad \ddot{x}+((q+1)\hat{k}_1x^q+(q+2)\hat{k}_2)\dot{x}
+(q+1)(\frac{(r-1)}{r^2}\hat{k}_1^2x^{2q}
\nonumber\\&&\qquad \qquad\qquad\qquad \qquad\qquad\qquad\qquad\qquad
+\hat{k}_1\hat{k}_2x^{q}+\hat{k}_2^2)x=0,\;\;r\neq0
\qquad \qquad \qquad\quad \nonumber (\ref {ncom6c})
\end{eqnarray}
where $k_1=(q+1)\hat{k}_1,\;k_2=(q+2)\hat{k}_2$.
Rewriting the associated first integral $I$, given in Case $(iv)$ in Table
$III$, in the form (\ref{the02}), we get
\begin{eqnarray}
I=\left\{
\begin{array}{ll}
\displaystyle{\bigg(\frac{(r-1)\hat{k}_1^2}{r^2}(xe^{\hat{k}_2t})^{2(q+1)}
+\frac {d}{dt}(xe^{\hat{k}_2t})\bigg(\frac {d}{dt}(xe^{\hat{k}_2 t})
e^{q\hat{k}_2t}+\hat{k}_1(xe^{\hat{k}_2t})^{q+1}\bigg)e^{q\hat{k}_2t}\bigg)},&\\\\
\qquad \qquad\times \displaystyle{\bigg(\frac {d}{dt}(xe^{\hat{k}_2t})e^{q\hat{k}_2t}
+\frac{\hat{k}_1(r-1)}{r}(xe^{\hat{k}_2t})^{q+1}\bigg)^{-r}},&
 r\neq0,2\\\\
\displaystyle{\frac{\frac {d}{dt}(xe^{\hat{k}_2t})e^{q\hat{k}_2t}}
{\frac{\hat{k}_1}{2}(xe^{\hat{k}_2t})^{q+1}+\frac {d}{dt}(xe^{\hat{k}_2t})
e^{q\hat{k}_2t}}-\log(\frac{\hat{k}_1}{2}(xe^{\hat{k}_2t})^{q+1}
+\frac {d}{dt}(xe^{\hat{k}_2t})e^{q\hat{k}_2t})},& r=2\\\\
\frac{1}{2}\bigg(\frac {d}{dt}(xe^{\hat{k}_2 t})\bigg)^2e^{2qk_2t}
+\frac{k_3}{2(q+1)}(xe^{\hat{k}_2 t})^{2(q+1)},& r=0.
\end{array}\right.\label{qthe20}
\end{eqnarray}
Identifying the dependent and independent variables from (\ref{qthe20}) 
and the relations (\ref{the03}), we obtain the transformation
\begin{eqnarray} 
w = xe^{\hat{k}_2 t}, \quad
z = -\frac {1}{q\hat{k}_2}e^{-q\hat{k}_2 t}.
\label{qthe21}
\end{eqnarray}
Subsituting the transformation (\ref{qthe21}) into (\ref{ncom6c}), one obtains 
\begin{eqnarray}
w''+(q+1)\hat{k}_1w^qw'+(q+1)\frac{(r-1)}{r^2}\hat{k}_1^2w^{2q+1}=0,\;\;r\neq0,
\quad '=\frac{d}{dz}.
\label{qthe24}
\end{eqnarray}
In terms of the new variables (\ref{qthe21}) change the time 
dependent first integral into time independent ones of the form
\begin{eqnarray}
I=\left\{
\begin{array}{ll}
\bigg(w'+\frac{(r-1)}{r}\hat{k}_1w^{q+1}\bigg)^{-r} 
\bigg[w'(w'+\hat{k}_1w^{q+1})+\frac{(r-1)}{r^2}\hat{k}_1^2w^{2(q+1)}\bigg],&
 r\neq0,2, \\\\
\frac{w'}{w'+\frac{\hat{k}_1}{2}x^{q+1}}-\log(w'+\frac{\hat{k}_1}{2}w^{q+1}),&
 r=2,\\\\
\frac{w'^2}{2}+\frac{k_3}{2(q+1)}w^{2(q+1)},&
 r=0.
 \end{array}\right. \label{qthe22}
\end{eqnarray}
Once again one can deduce the Hamiltonians in the form
\begin{eqnarray}
H=\left\{
\begin{array}{ll}
\bigg[\frac{\bigg((r-1)p\bigg)^{\frac{r-2}{r-1}}}{(r-2)}-
\frac{(r-1)}{r}\hat{k}_1w^{q+1}p\bigg],& r\neq0,1,2,\\\\
\frac{1}{2}\hat{k}_1w^{q+1}p+\log(\frac{2(q+1)}{p}),& r=2,\\\\
e^p+\hat{k}_1w^{q+1},& r=1,\\\\ 
\frac{p^2}{2}+\frac{k_3}{2(q+1)}w^{2(q+1)},r=0,
\end{array}\right. \label{qthe22e}
\end{eqnarray}
with
\begin{eqnarray}
p=\left\{
\begin{array}{ll}
\frac{1}{(r-1)}\bigg(w'+\frac{(r-1)}{r}\hat{k}_1w^{q+1}\bigg)^{(1-r)},&
r\neq0,1\\\\
\log(w'),&r=1\\\\ 
w',& r=0,
\end{array}\right. \label{qthe22ee}
\end{eqnarray}
and thereby ensuring liouville integrability of Eq.~(\ref{ncom6c}).
\subsection{Summary of results in $q=\;arbitrary$ case:}
\label{sec53}
To conclude the integrability of Eq.~(\ref{int08}), we have established the fact
that the following equations, are integrable
\begin{eqnarray}
 (1) \quad &&\ddot{x}+(k_1x^q+(q+2)k_2)\dot{x}+k_1k_2x^{q+1}
+(q+1)k_2^2 x=0,
\qquad \qquad \qquad \qquad \qquad\;
\;\;\;\; \nonumber (\ref {ncom11a})\\        
 (2) \quad&& \ddot{x}+((q+2)k_1x^q+k_2)\dot{x}+k_1^2x^{2q+1}
+k_1k_2x^{q+1}+\lambda_1x=0, \qquad \qquad \qquad \qquad \;\;\;\;\;\;\;  \nonumber 
(\ref {ncom7})\\
 (3) \quad&&\ddot{x}+(q+4)k_2\dot{x}+k_4x^{q+1}+2(q+2)k_2^2x=0,
\qquad \qquad \qquad\qquad\qquad \qquad\qquad\;\;\;\;\;\; \nonumber (\ref {ncom11})\\
 (4)\quad&& \ddot{x}+((q+1)k_1x^q+k_2)\dot{x}+\frac{(r-1)}{r^2}((q+1)k_1^2x^{2q}
\nonumber\\&&\qquad \qquad\qquad \qquad\qquad \qquad
+(q+2)k_1k_2x^{q}+k_2^2)x=0,\;r\neq0
\qquad\qquad \qquad \qquad  \nonumber (\ref {ncom9})\\  
 (5) \quad&&\ddot{x}+((q+1)k_1x^q+(q+2)k_2)\dot{x}
+(q+1)(k_3x^{2q}+k_1k_2x^{q}+k_2^2)x=0,
\qquad \qquad \quad \nonumber (\ref {ncom6c})
\end{eqnarray}
where $r^2k_3=(r-1)k_1^2$ and $k_1,\;k_2,\;k_4,\;\lambda_1$ and $r$ are 
arbitrary parameters (for simplicity we have removed hats in $k_i$'s,
$i=1,2,$ in Eqs.~(\ref{ncom11a})-(\ref{ncom6c})) .
The significance and newness of the equations~(\ref{ncom11a})-(\ref{ncom6c}) are
already pointed out in Sec. \ref{sec12}.

\section{Discussion and Conclusions}
\label{sec6}
In this paper, we have investigated the integrability properties of
Eq.~(\ref{int08}) and shown that it admits a large class
of integrable nonlinear systems. In fact, many classical integrable 
nonlinear oscillators  can be derived as sub-cases of our results. One of the 
important outcomes of our investigation is that the entire 
class of Eq.~(\ref{int05}) can be
derived from a conservative Hamiltonian (vide Eq.~(\ref{qlin012})) eventhough 
the system deceptively looks like a dissipative equation.

From our detailed analysis we have shown that Eq.~(\ref{int08}) admits both
conservative Hamiltonian systems and dissipative systems, depending on the
choice of parameters. As far as the former is
concerned we have deduced the explicit forms of the Hamiltonians for
the respective equations. In fact, for the case, $q=1$, we have constructed
suitable
canonical transformations and transformed the equations into conservative
nonlinear oscillator
equations. However, the canonical transformations for the conservative Hamiltonian
systems for the cases $q=2,\ldots,$ arbitrary, if at all they exist, still remain
to be obtained. Exploring the classical dynamics underlying these conservative
Hamiltonian systems is also of consideroble interest for further study.
As far as dissipative systems are concerned we have not only shown that
Eq.~(\ref{int08}) contains the well known force-free Helmholtz, Duffing and  
Duffing-van der
Pol oscillators but also have several integrable generalizations which is
another important outcome of our investigations. The study of chaotic dynamics
of these nonlinear oscillators under further perturbations is one of the 
current topics$\footnotesize^{22}$ in the contemporary
literature in nonlinear dynamics. In principle one can extend such analysis
to the above generalized equations as well.

In this paper, we have also not touched the question of linearizability of the
integrable cases of Eq.~(\ref{int08}). In our earlier work, we have shown that the
Eq.~(\ref{case01}) is linearizable to the free particle equation,
$\frac{d^2w}{dz^2}=0$. Of course one can show that this is the only linearizable
equation in (\ref{int08}) through invertible point 
transformation$\footnotesize^{9,11,18}$. However, linearizablity of other
integrable cases through more general transformations still remains to be
explored.

In addition to the above, we have also carried out the Painlev\'e singularity 
structure  analysis of 
Eq.~(\ref{int08}) and compared the results obtained through both the methods. 
The details of this will be published elsewhere.

As we mentioned at the end of Sec. \ref{sec2}, the crux of the PS procedure lies in
finding the explicit solutions satisfying all the three determining
Eqs.~(\ref{lin02})-(\ref{lin04}). In this paper we have considered only certain specific
ansatz forms to determine the null forms $S$, and integrating factors $R$. As a
consequence only a specific class of integrable equations have been derived. It
is not clear, whether these ansatz forms used in this paper exhaust all possible
integrable cases of Eq.~(\ref{int08}). One needs to consider
more generalized ansatz forms, and if possible to solve 
Eqs.~(\ref{lin02})-(\ref{lin04})
for the most general forms of $R$ and $S$, and try to identify all possible
integrable cases underlying Eq.~(\ref{int08}). This is being explored further.

\section{Acknowledgments}
The work of VKC is supported by CSIR in the form of a Senior Research
Fellowship.  The work of MS and ML forms part of a Department of Science 
and Technology, Government of India, sponsored research project.

\begin{tabular}{p{.15cm}p{14cm}}
\footnotesize$^1$ &
V. K. Chandrasekar, M. Senthilvelan and M. Lakshmanan,  J. Phys. A 
{\bf 37}, 4527 (2004).\\
\footnotesize$^2$ &
S. N. Pandey, M. Senthilvelan and M. Lakshmanan, {\it Classification of Lie point
symmetries of nonlinear dissipative system of the type
$\ddot{x}+f(x)\dot{x}+g(x)=0$} preprint, to be submitted for publication. \\
\footnotesize$^3$ &
M. R. Feix, C. Geronimi, L. Cairo, P. G. L. Leach, R. L. Lemmer  and S. 
Bouquet, J. Phys. A {\bf 30}, 7437 (1997).\\
\footnotesize$^4$ &
R. A. Smith, J. London Math. Soc. {\bf 36}, 33 (1961).\\
\footnotesize$^5$ &
M. Prelle and M. Singer, Trans. Am. Math. Soc. {\bf 279}, 215 (1983).\\
\footnotesize$^6$ &
Y. K. Man and M. A. H. MacCallum, J. Symbolic Computation {\bf 11}, 1
(1996).\\
\footnotesize$^7$ &
L. G. S. Duarte, S. E. S. Duarte, A. C. P. da Mota and J. E. F. Skea,  
J. Phys. A {\bf 34}, 3015 (2001).\\
\footnotesize$^8$ &
L. G. S. Duarte, S. E. S. Duarte and A. C. P. da Mota, J. Phys. A 
{\bf 35}, 1001 (2002); {\bf 35}, 3899 (2002).\\
\footnotesize$^{9}$ &
V. K. Chandrasekar, M. Senthilvelan and M. Lakshmanan, 
Proc. R. Soc. London {\bf A461}, 2451 (2005).\\
\footnotesize$^{10}$ &
V. K. Chandrasekar, M. Senthilvelan and M. Lakshmanan, J. Nonlinear 
Math. Phys. {\bf 12}, 184 (2005). \\
\footnotesize$^{11}$ &
V. K. Chandrasekar, M. Senthilvelan and M. Lakshmanan, Chaos, Solitons 
and Fractals {\bf 26}, 1399 (2005).\\ 
\footnotesize$^{12}$ &
M. Lakshmanan and S. Rajasekar, { \it Nonlinear Dynamics: Integrability
Chaos and Patterns} (Springer-Verlag, New York, 2003).\\
\footnotesize$^{13}$ &
J. A. Almendral and M. A. F. Sanju\'an, J. Phys. A
{\bf 36}, 695 (2003).\\
\footnotesize$^{14}$ &
S. Parthasarathy and M. Lakshmanan, J. Sound and Vib. {\bf 137}, 523 (1990).\\ 
\footnotesize$^{15}$ &
G. W. Bluman and S. C. Anco, {\it Symmetries and Integration Methods
for Differential Equations} (Springer-Verlag, New York, 2002).\\
\footnotesize$^{16}$ & 
T. C. Bountis, L. B. Drossos, M. Lakshmanan and S. Parthasarathy 
J. Phys. A {\bf 26}, 6927 (1993).\\
\footnotesize$^{17}$ &
V. K. Chandrasekar, M. Senthilvelan and M. Lakshmanan,  {\it On the complete 
integrability and linearization of nonlinear ordinary differential equations-
part II: third order equations} submitted to Proc. R. Soc. London A.\\
\end{tabular}

\begin{tabular}{p{.15cm}p{14cm}}

\footnotesize$^{18}$ &
F. M. Mahomed and P. G. L.  Leach, Quaestiones Math. {\bf 8}, 241 (1985); 
{\bf 12} 121 (1985); P. G. L. Leach, M. R. Feix and S. Bouquet, 
J. Phys. A {\bf 29}, 2563 (1988); P. G. L. Leach, 
 J. Math. Phys. {\bf 26}, 2510 (1985); R. L. Lemmer and P. G. L. Leach, 
J. Phys. A {\bf 26}, 5017 (1993). \\
\footnotesize$^{19}$ &
V. K. Chandrasekar, M. Senthilvelan and M. Lakshmanan, 
arXiv:nlin. SI/0408054 submitted to Phys. Rev. E. \\
\footnotesize$^{20}$ &
E. L. Ince, {\it Ordinary Differential Equations} (Dover, New York, 1956).\\
\footnotesize$^{21}$ &
G. M. Murphy, {\it Ordinary Differential Equations and Their Solutions} 
(Affiliated East-west press, New Delhi, 1969).\\
\footnotesize$^{22}$ &
D. L. Gonzalez and O. Piro, Phys. Rev. Lett. {\bf 50}, 870 (1983); 
Phys. Rev. {\bf A30}, 2788 (1983).\\
\footnotesize$^{23}$ &
M. Lakshmanan and J. Prabhakaran, 1973 Lettere al Nuovo Cimento {\bf 7}, 
689 (1973); M. Lakshmanan, Lettere al Nuovo Cimento {\bf 8}, 743 (1973).\\
\footnotesize$^{24}$ &
I. S. Gradshteyn and I. M. Ryzhik, {\it Table of Integrals, Series and 
Products} (Academic press, London, 1980).\\

\end{tabular}

\end{document}